\documentclass[a4paper,twocolumn]{article}
\pdfoutput=1
\usepackage{amsmath}
\usepackage{amsfonts}
\usepackage{amssymb}
\usepackage{graphicx}
\usepackage{natbib}
\usepackage{authblk}
\usepackage{color}
\usepackage{balance}
\usepackage{titlesec}
\usepackage{booktabs}
\usepackage{tabularx}
\usepackage{multirow}
\usepackage{abstract}
\usepackage[font=footnotesize,labelfont=bf]{caption}

\titleformat*{\subsubsection}{\it}

\newcommand{\msection}[1]{\section*{\centering#1}}\newcommand{\textsim}{$\sim$}

\newcommand{\Nrem}{$N_{rem}$}
\newcommand{\Ndep}{$N_{dep}$}
\newcommand{\vimp}{$v_{imp}$}
\newcommand{\rimp}{$r_{imp}$}
\newcommand{\rt}{$r_{t}$}
\newcommand{\dimp}{$d_{imp}$}
\newcommand{\Nimp}{$N_{imp}$}
\newcommand{\rmin}{$r_{min}$}
\newcommand{\rmax}{$r_{max}$}
\newcommand{\rand}{\mathcal{R}}
\newcommand{\qstard}{$Q_D^{\star}$}
\newcommand{\qdqstard}{$Q/Q_D^{\star}$}
\newcommand{\rstari}{$r_i^{\star}$}
\newcommand{\qheat}{$Q_{heat}$}
\newcommand{\degr}{$^{\circ}$}
\newcommand{\etan}[1]{$\eta_{#1 r_t}$}
\newcommand{\Df}{$D_f$}
\newcommand{\Qtot}{$Q_{tot}$}
\newcommand{\Qheat}{$Q_{heat}$}
\newcommand{\Hpen}{$H_{pen}$}
\newcommand{\dtr}{$d_{tr}$}

\title{\bf{The early impact histories of meteorite parent bodies}}

\author[1,2]{Thomas M. DAVISON \thanks{e-mail address: thomas.davison@imperial.ac.uk}}
\author[3]{David P. O'BRIEN}
\author[1]{Fred J. CIESLA}
\author[2]{Gareth S. COLLINS}

\affil[1]{\small{Department of the Geophysical Sciences, University of Chicago, 5734 S Ellis Avenue, Chicago, IL 60637, USA}}
\affil[2]{\small{Department of Earth Science and Engineering, Imperial College London, London, SW7 2AZ, UK}}
\affil[3]{\small{Planetary Science Institute, 1700 E. Ft. Lowell, Suite 106, Tucson, AZ 85719, USA}}

\date{\footnotesize To be published in Meteoritics \& Planetary Science \\
DOI:10.1111/maps.12193}

\begin{document}
\twocolumn[
  \maketitle
  \vskip-80pt
  \begin{onecolabstract}
We have developed a statistical framework that uses collisional evolution models, shock physics modeling and scaling laws to determine the range of plausible collisional histories for individual meteorite parent bodies.  It is likely that those parent bodies that were not catastrophically disrupted sustained hundreds of impacts on their surfaces Ñ compacting, heating, and mixing the outer layers; it is highly unlikely that many parent bodies escaped without any impacts processing the outer few kilometers.  The first 10 -- 20 Myr were the most important time for impacts, both in terms of the number of impacts and the increase of specific internal energy due to impacts.  The model has been applied to evaluate the proposed impact histories of several meteorite parent bodies: Up to 10 parent bodies that were not disrupted in the first 100 Myr experienced a vaporizing collision of the type necessary to produce the metal inclusions and chondrules on the CB chondrite parent; around 1 -- 5\% of bodies that were catastrophically disrupted after 12 Myr sustained impacts at times that match the heating events recorded on the IAB/winonaite parent body; more than 75\% of 100 km radius parent bodies which survived past 100 Myr without being disrupted sustained an impact that excavates to the depth required for mixing in the outer layers of the H chondrite parent body; and to protect the magnetic field on the CV chondrite parent body, the crust would have had to have been thick (\textsim\ 20 km) in order to prevent it being punctured by impacts.\\

  \end{onecolabstract}
\strut
]
\saythanks

\msection{INTRODUCTION}
The early Solar System was a violent place for young planetesimals.  Collisions with other planetesimals were common, and would have affected the evolution of the bodies that would go on to become the asteroids and parent bodies of the meteorites we find on Earth today.
Planetesimals, the first rocky bodies to form in the Solar System, accreted within the solar nebula, the cloud of dust and gas that orbited the young Sun.  Initially, highly porous dust aggregates up to centimeters in scale accreted in low-velocity, pairwise collisions between dust particles \citep{Blum:03,Wurm:01,Wurm:04}.  How these aggregates grew from this scale to the kilometer scale is still an open question, although leading theories suggest that either hydrodynamic effects and gravitational instabilities \citep{Johansen:07, Johansen:09, Cuzzi:08} or further low velocity collisions led to a population of planetesimals \textsim~1 -- 100 km in scale \citep{Weidenschilling:11}.  These early bodies are likely to have contained a significant fraction of pore space \citep{Dominik:97, Bland:11}.  Some planetesimals grew to protoplanets; the first stage of this growth is termed runaway growth \citep{Greenberg:78, Wetherill:89, Wetherill:93}, in which the largest planetesimals gravitationally focused smaller bodies, increasing the number of collisions on their surfaces \citep{Kokubo:96}, eventually drawing in all bodies from their so-called Ôfeeding zoneÕ.  As the number of small bodies was diminished during runaway growth, protoplanets grew more slowly during the subsequent oligarchic growth phase, as impactors needed to be drawn from further afield and the protoplanets began to excite the remaining planetesimal population, reducing the number of collisions on their surfaces \citep{Kokubo:98}. 

These processes are expected to have resulted in a remnant population of planetesimals and a smaller number of embryos (\textsim~100~--~1000~km in radius).  Collisions between and among these populations would have been common, as the number of bodies would have been much greater than the number of bodies that remain in the asteroid belt today. For example, the population in the region 2 -- 4 AU could have been \textsim~100~--~1000 times more massive than the current asteroid belt \citep{Bottke:05b,Bottke:05a}.  Mass was lost from the population as bodies were scattered into the Sun or ejected out of the Solar System by Jupiter; the majority of this mass depletion took place in the initial 100 Myr period \citep{Petit:01,Bottke:05b}, meaning that the frequency of collisions during that period was much greater than it was after mass depletion took place.  Planetary embryos could have `stirred up' the smaller planetesimal population, increasing velocities above the mutual escape velocity of the colliding pair of bodies, up to several kilometers per second \citep{Kenyon:01}.  As some bodies fell into resonances with Jupiter and Saturn, their orbits would have been excited to high eccentricities Ñ leading to impact velocities of at least 10 km/s \citep{Weidenschilling:98, Weidenschilling:01, Bottke:05b}.

As meteorites provide our strongest evidence of conditions in the early Solar System, a full understanding of the histories of their parent bodies is vital.  Collisions played a major role in these histories:  Indeed, impacts have been invoked to explain many observations from the meteorite record.  For example, shock wave processes have been shown to cause deformation of minerals and localized melting \citep{Stoffler:91} and shock blackening \citep{Heymann:67, Britt:91};  \citet{Wittmann:10} show that these effects typically postdate metamorphism from radiogenic decay.  Impacts are able to perturb the Ôonion-shellÕ nature of petrologic types; for example, \citet{Grimm:85} and \citet{Taylor:87} proposed nearly complete mixing of chondrite parent bodies by impact disruption and reaccumulation. More recently, it has been suggested that impact processing may have brought some heated material (petrologic type 4--6) from the interior of the parent body into close contact with cooler material (type 3) from the outer layers of the body \citep{Harrison:10a,Scott:11a}.  Impacts have even been invoked as a possible heating mechanism leading to resetting of thermochronometers or phase transitions in more extreme cases, for example on the IAB/winonaite parent body \citep{Schulz:09, Wasson:02a}, the CB chondrites \citep{Campbell:02b, Krot:05} and the Angrite and Eucrite parent bodies \citep{Scott:11b}.  Recently the effects of individual collisions between porous bodies has been quantified \citep{Davison:10a}, as well as their influence on the long timescale thermal evolution of a parent body \citep{Davison:12}, demonstrating that localized heating can be significant and allow higher temperatures to be reached than previously estimated assuming non-porous bodies \citep{Keil:97}.

The extent to which impacts would have affected a meteorite parent body will vary from body to body, and depend on the number, sizes and timing of the collision events.  To date, there has been no quantitative analysis of the types of collision histories that a meteorite parent body might have experienced during its early evolution.  Analytical techniques \citep[e.g][]{Chambers:06}, can provide estimates of the number of mutual collisions between planetesimals of a given size. However, this approach does not provide histories for individual parent bodies that are subject to impacts from an evolving population of smaller planetesimals.  To investigate the role of collisions on the thermal and disruptional histories of meteorite parent bodies, and to find the likelihood that collisions could have produced the effects they have been invoked to explain in the meteorite record, it is important to know what kinds of collisions occurred.

This study aims to address this gap in our knowledge of impact processes, and quantify: (1) how many impacts a parent body is likely to sustain during the first ~ 100 Myr of its history; (2) how likely a parent body is to be disrupted during this time; (3) when collisions are most likely to occur; (4) the average and range of velocities of the impacts and how they change with time; (5) the time that most disruptive impacts occur; and (6) how much collisional heating occurs during that time.
In this paper we develop a Monte Carlo model to determine the history of impacts on a parent body surface, and use this model to provide answers to the above questions.  In the next section, the details of the model are presented.  Then, results from this model for a range of parent bodies and collisional evolution scenarios are presented.  Finally, the predictions from the model are applied to several case studies for which impacts have been invoked in the literature, to test the plausibility of those scenarios.

\msection{METHODS}
A large number of planetesimals would have been present in the early Solar System, each with their own collisional history determined by a series of chance encounters with other planetesimals.  Therefore the impact history of a parent body through time cannot be solved analytically, as each body would have experienced a different set of impacts on its surface.  A Monte Carlo approach is required to determine the range of possible impact histories that a parent body could have experienced, and what a typical impact history would have been.

In the Monte Carlo simulation developed for this study, the collisional histories of many parent bodies are modeled, where the likelihood of a collision with a given impactor size occurring at a given time and the velocity of each collision are calculated from a probability distribution.  The probability of a collision occurring with a given impactor size and the velocity-frequency distribution at a time t must therefore be known beforehand.  These distributions are based on dynamical and collisional models of terrestrial planet formation; their calculations are outlined below, in the section ÔCollisional evolution modelingÕ.  For each collision, the resulting crater size is estimated from crater scaling relationships, the amount of heating is determined from shock physics simulations, and whether the parent body is catastrophically disrupted can be calculated using strength estimates from simulations of catastrophic collisions \citep{Benz:99, Jutzi:10}.  These calculations are outlined in the section `\emph{Collisions: Cratering and disruption}'.

\subsection*{Collisional and dynamical evolution modeling}
Quantifying the impact history of a parent body requires knowledge of the size-frequency distribution (SFD) of planetesimals, the velocity-frequency distribution (VFD) of collisions and the intrinsic collisional probability, which are all known to be time-dependent.  Simulations of the dynamical and collisional evolution of the planetesimal population in the early Solar System are able to provide estimates of these quantities. \citet{O'Brien:06, O'Brien:07}
OÕBrien et al. (2006; 2007) performed simulations of terrestrial planet formation and the excitation and mass depletion of the early asteroid belt using the SyMBA N-Body integrator \citep{Duncan:98}.  The gravitational interactions of a population of Mars-mass planetary embryos and an equal mass of smaller planetesimals distributed from 0.3--4 AU were simulated, including the influence of Jupiter and Saturn.  The smaller planetesimal bodies are influenced by the gravity of the large bodies, but do not interact with each other, as is commonly done in such simulations, while the larger bodies are influenced by the gravity of the small planetesimals and one another.  Collisions occur when an embryo runs into another embryo or planetesimal, which leads to a perfect merger of the impacting bodies, conserving linear momentum.  

While mutual collisions between planetesimals were ignored in the planetary accretion simulations, they are expected to have occurred during this period of planet formation.  As such, the simulations were analyzed to determine the evolution of the collision probability and VFD for the first 100~Myr of dynamical evolution using the algorithm of \citet{Bottke:94}, where $t = 0$ is defined as the time when most of the gas has dissipated and gas drag no longer provides a damping effect. Bodies were divided into two groups: Those that remained in the asteroid belt at the end of the simulations, and those that were either accreted by the terrestrial planets or lost from the system.  Collision velocity distributions and intrinsic collision probabilities, $P_i$, were calculated for planetesimals within and between these different populations throughout the time of interest (Figure 1).

\begin{figure}[t]
\includegraphics[width=0.9\linewidth]{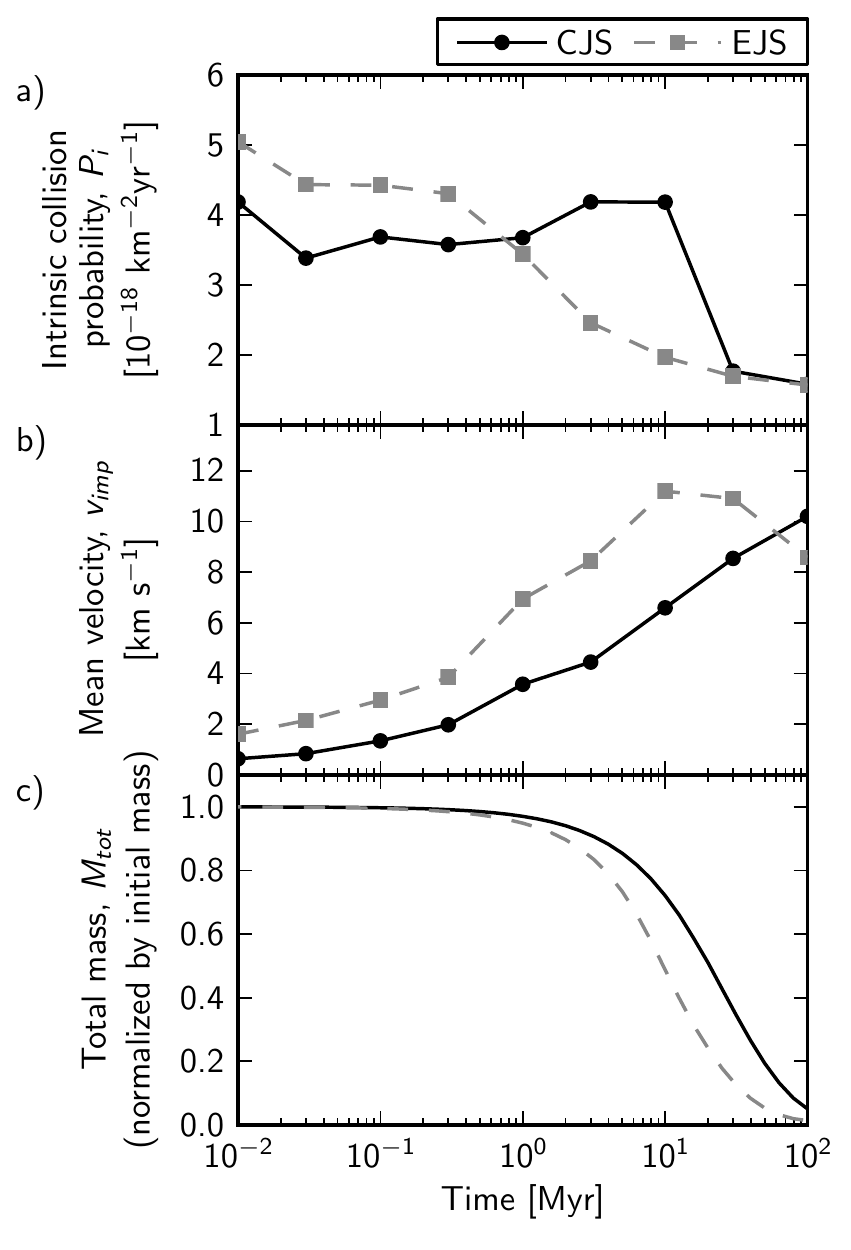}
\caption{Evolution of the intrinsic collisional probability, the mean collision velocity and the total mass of planetesimals for the two different collisional evolution models described in the text, CJS and EJS. Collisional probability and velocity are specifically for collisions of all planetesimals onto the planetesimals that remain in the asteroid belt at the end of the simulations.}
\end{figure}

The dynamical simulations just consider a single particle size for the planetesimals, so to track the evolution of the size distribution, we use a separate collisional evolution model that takes the time-dependent impact velocity, collision probabilities, and total mass determined from the dynamical simulations and evolves the size distributions of the planetesimal populations in time under mutual collisions, assuming an impact strength for porous planetesimals \citep{Jutzi:10}.

This model is modified from \citet{O'Brien:05} and \citet{O'Brien:09} to treat multiple interacting populations and time-dependent collisional parameters, in an approach similar to that of \citet{Bottke:05b}.  Two separate but interacting populations are considered: The bodies that eventually end up in the asteroid belt region (denoted \Nrem), for the remnant population), and all other bodies (those that are eventually lost and those that are accreted by the terrestrial planets, as described above; denoted \Ndep for the ÔdepletedÕ population).  The size of the depleted population is varied to provide a good match between the final remnant population in the model and the observed mass of the present-day asteroid belt, where \Ndep$ = f $\Nrem.  For the non-asteroid belt bodies, a forced exponential decay term is applied, based on how fast the populations deplete in the dynamical simulations.  As time advances, the production of collisional fragments increases the number of small bodies in both populations (Figure 2).  Both populations evolve through mutual collisions, but eventually (after tens to \textsim~100~Myr) the non-asteroid belt bodies are dynamically depleted and their contribution to the collisional evolution of the asteroids is negligible, leaving only the asteroids.  This method is discussed in more detail in \citet{Bottke:05b}.
 
\begin{figure}[!t]
\includegraphics[width=0.9\linewidth]{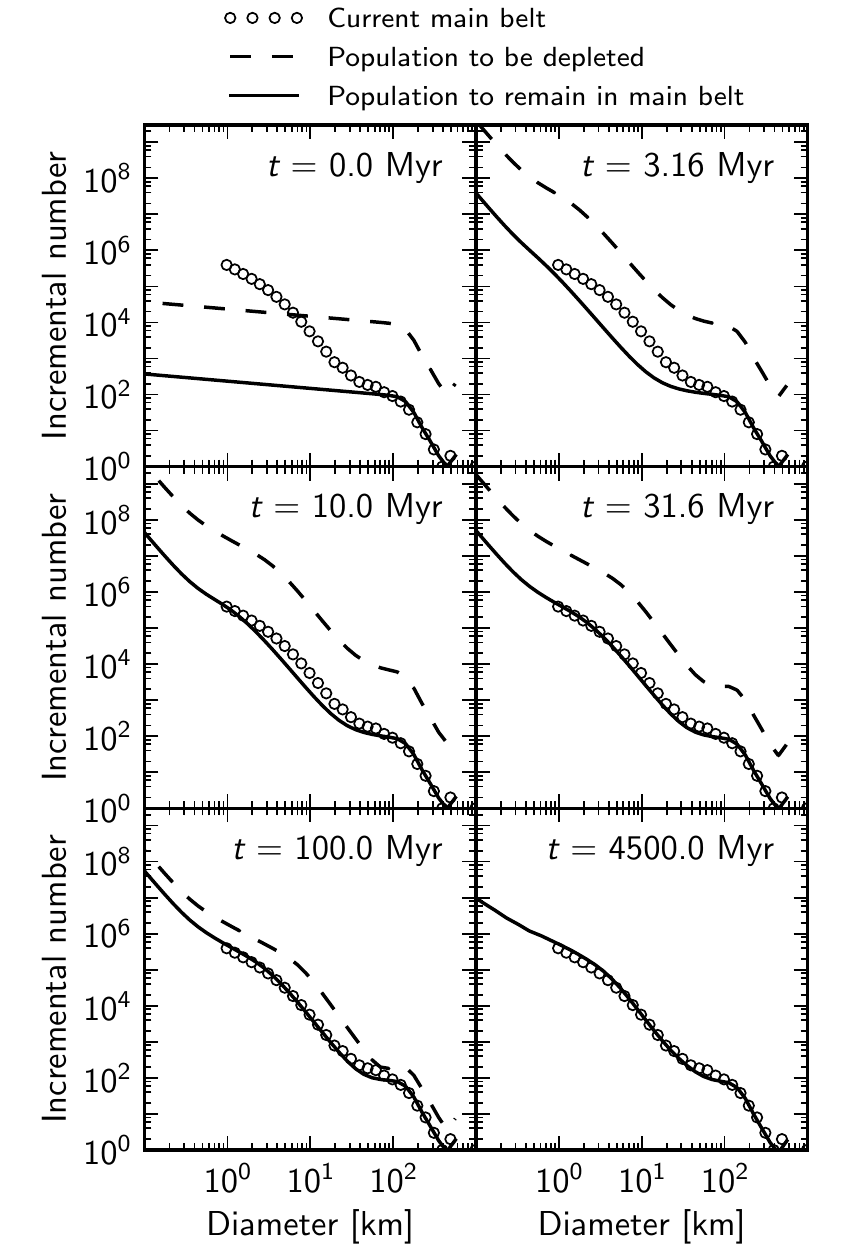}
\caption{Evolution of the two planetesimal populations over the course of the CJS simulation with $f = 100$. The number of small bodies rapidly increases from the initial distribution due to collisional fragmentation. Depletion of the population begins at $t = 0$, and decays with a half-life of \textsim~22 Myr (for comparison, the EJS model decays with a half-life of \textsim~11 Myr). After 4500 Myr, the depleted population has been completely removed from the disk, leaving only the bodies that remain in the asteroid belt.}
\end{figure}
 
 These dynamical and collisional evolution simulations depend upon the orbits of the gas giants Jupiter and Saturn.  The initial orbital parameters of the gas giants are not well known.  In this work we consider two different end-member simulations, which were found to give reasonable agreement between the final SFD of planetesimals and the current SFD of the asteroid belt.  In both simulations, Jupiter and Saturn were included from the start of the simulation \citep{Chambers:01, O'Brien:06}. In one set of simulations, Jupiter and Saturn started with orbits consistent with their current inclination and eccentricity (denoted EJS here, for `Eccentric Jupiter and Saturn').  The embryos and planetesimals were centered on the invariant plane of Jupiter and Saturn.  In this simulation, the final mass of the planetesimals in the asteroid belt region was a factor of 95 less than the total initial mass of the planetesimals in the system (i.e. $f = 95$).  In the other set of simulations, Jupiter and Saturn adopted initial orbits that were near-circular and co-planar (i.e. inclination and eccentricity were essentially zero), similar to that expected based on the Nice model \citep[denoted CJS here, for `Circular Jupiter and Saturn']{Tsiganis:05, Morbidelli:05, Gomes:05} denoted CJS here, for ÔCircular Jupiter and SaturnÕ).  Planetesimals and embryos were centered on the plane of JupiterÕs initial orbit.  In these simulations, the best match to the final SFD of the asteroid population was found when the final mass of the planetesimals in the asteroid belt region was a factor of 100 less than the total initial mass of the planetesimals in the system (i.e. $f = 100$, which is similar to the $f$ value found for the EJS simulation and the results of \citet{Bottke:05b}, who used $f$ values in the range \textsim~100~--~200, depending on the formation time of Jupiter).  Figure 2 shows the evolution of the remnant and depleted populations in the CJS simulation with $f = 100$. The fit here is in close agreement with the observed SFD of the current asteroid belt. However, the distribution has fewer bodies between a few tens of km and 100~km in diameter than the current asteroid belt, which may overestimate the collisional grinding in this size-range (i.e. too many bodies of that size have experienced a disruptive collision), which in turn produces a larger number of small objects at the expense of larger ones. 
 
Figure 1 shows the evolution through time of the intrinsic collisional probability, $P_i$ (Figure 1a in units of km$^{-2}$ yr$^{-1}$; see Equation (1) to convert $P_i$ to a probability of a collision between a pair of planetesimals of radii \rimp~and \rt, in a given time interval), the mean collision velocity (Figure 1b) and the total mass of planetesimals (Figure 1c), for the two different simulations.  For the EJS simulation, impact velocities increase sooner than for CJS, and the collisional probability is higher in early times.  However, after \textsim~1~Myr, the collisional probability in the EJS case decreases, while for CJS it remains approximately constant until about 10~Myr until dropping off.  This is because in the EJS case bodies are rapidly excited to high inclination orbits, leading to the depletion of objects from the simulation, and thus fewer collisions occur after \textsim~1~Myr.  In the CJS case, this excitation takes longer, and therefore the collisional probability does not decrease as quickly.  These differences could play an important role in the collisional evolution of a parent body, and the sensitivity of our results to these assumptions is tested below.

The amount of damage done to main belt asteroids in the first few 100~Myr is approximately equivalent to that done in the proceeding 4.4~Gyr, suggesting that the latter component probably cannot be ignored.  However, the big difference in terms of impact heating is that (i) the parent bodies are now cold and (ii) impact velocities are now much lower (most impacts come from main belt projectiles), such that the amount of heat produced will be much lower.  Thus, for applications of impact heating, or discussions of the early impact histories of parent bodies, this model is applicable.  If we wish to apply the results of this model to study the complete impact history of a parent body, the model would need to extend over the full history of the Solar System.
It should be noted that the model employed here does not account for radial migration of giant planets, such as that suggested in the Grand Tack model \citep{Walsh:11}, and thus that scenario is not treated in this work.  The approaches outlined here, however, are general and we plan to investigate that particular model of Solar System formation in a future study.

\subsection*{Monte Carlo Model}
In this section, the model algorithm used in this work is described.  For each parent body in our model (one sampling of the Monte Carlo model), we step through time from $t = 0$ to $t = 100$~Myr.  At each time, $t$, the probability, $P_c$, that a collision will occur between the parent body, of radius \rt, and an impactor of size \rimp\ at this time, $t$, is calculated:

\begin{equation}
P_c\left(r_{imp},t\right)=N_{imp}P_i\left(r_{imp}+r_{t}\right)^2\Delta t
\end{equation}

The intrinsic collision probability, $P_i$, and number of impactors in a given size bin, \Nimp, at a given time $t$, were estimated by interpolating between data points output from the dynamical models described above, where the SFD of bodies in the collisional evolution models represented the population of potential impactors. While the collisional evolution models provide statistics for the number of planetesimals with radii, \rimp, ranging from \textsim~1~m to 500~km,  to increase calculation speeds, and because smaller impactors are unable to disrupt or significantly process the parent body, only impactors with \rimp~$>$~150~m are considered.   The boundaries of the impactor size bins are logarithmic, with $\log_{10}\left(r_{\rm{max}}/r_{\rm{min}}\right)=0.1$, where \rmax\ is the upper bound of the radius of a planetesimal in a given size bin, while \rmin\ is the lower bound.  If t is between two of the output timesteps from the collisional evolution modeling, \Nimp\ was determined using a linear interpolation on the SFD.

Next, a random number, $\rand_1$, was generated, where 0 $\leq \rand_1 \leq 1$. If $\rand_1 < P_c\left(r_{\rm{imp}},t\right)$, then a collision was adjudged to occur between an impactor of size \rimp\ and the parent body at this time, $t$.  If a collision occurs, then the collateral effects of that collision are quantified: A second random number, $\rand_2$, is chosen, to pick a collision velocity from the Maxwellian distribution of velocity (defined by the mean velocity, $\bar{v}$), which was also interpolated as needed (see appendix).

\subsubsection*{Collisions: Cratering and disruption}
With knowledge of the impactor size and velocity, and the parent bodyÕs size (and hence gravity), the diameter of the transient crater formed by the impact, $d_{tr}$, can be determined from crater scaling relationships determined from experiments \citep{Schmidt:87, Holsapple:93} and numerical modeling \citep{O'Keefe:93, Pierazzo:97}:

\begin{equation}
d_{tr}=1.61^{-\beta}C_Dg^{-\beta}d_{imp}^{1-\beta}v_{imp}^{2\beta}\left(\dfrac{\pi}{6}\right)^{\frac{1}{3}}
\end{equation}

where g is the surface gravity of the parent body, \dimp\ is diameter of the impactor (= 2\rimp), and \vimp\ is the impact velocity. The material-specific constants for non-porous soil and sand were estimated by \citet{Schmidt:87} as $C_D$ = 1.6, $\beta$ = 0.22 and $C_D$ = 1.54, $\beta$ = 0.165, respectively. Based on numerical cratering simulations, \citet{Wunnemann:06} suggest that the constants for sand represent a good approximation for moderately porous, granular materials.

To convert the transient crater diameter to a final crater diameter, it is necessary to know whether the crater is a simple, bowl-shaped crater or a complex crater. The transition from simple to complex craters is known to occur at \textsim~3.2~km on Earth, and \textsim~18~km on the Moon, and scales as the inverse power of the surface gravity, $g$. Indeed, this relationship is thought to extend over a wide range of surface gravities, including to small bodies such as Vesta \citep{Melosh:99}. Hence, the simple-to-complex transition diameter, $d_{sc}$, can be estimated by:

\begin{equation}
d_{sc}=\dfrac{g_{moon}\rho_{moon}d_{sc_{moon}}}{g\rho_t}
\end{equation}

where $g_{moon}$, $\rho_{moon}$ and $d_{sc}$ are the surface gravity, density and simple-to-complex transition diameter on the moon, respectively, and $\rho_t$ is the density of the parent body.  The final crater diameter can then be estimated, depending on whether it is simple or complex, by the following relationships \citep{McKinnon:85, Collins:05}.  For a simple crater, the final diameter, $d_f$, is given by:

\begin{equation}
d_f=1.25d_{tr}	
\end{equation}

and the final crater diameter for a complex crater is given by:

\begin{equation}
d_f=1.17\dfrac{d_{tr}^{1.13}}{d_{sc}^{1.13}}
\end{equation}

Equations (2 -- 5) assume that all craters occur in the gravity regime.  Estimates of the transition between strength-dominated craters and gravity-dominated craters suggest that on a 100 km radius body, any crater larger than ~ 2.3 km will be in the gravity regime \citep{Nolan:96}.  As the minimum size of impactors considered in this study is \textsim~150~m radius, only a small proportion of craters are in the strength regime.  For the simulations presented below, less than 0.1\% of impacts are in the strength regime for 100 km and 250 km radius parent bodies.  For the simulations of 50 km radius parent bodies, around 2 -- 3\% of craters are in the strength regime.  As these are also the smallest craters, they have little effect on the overall statistics of the size of craters on the parent body.  If future studies are conducted to investigate the collisional histories of smaller parent bodies, the strength regime should be accounted for in the calculation of crater size.
      
The depth that the impact event excavates material from (i.e. the maximum depth from which material can be brought to the surface, rather than being displaced downwards into the target) can also be determined from scaling laws and the same impactor parameters (impactor size and velocity, and target gravity). \citet{Melosh:89} states that the excavation depth, $H_{exc}$, is approximately one tenth of the transient crater diameter, i.e.:
      
\begin{equation}
H_{exc} = \dfrac{1}{10}d_{tr}
\end{equation}

One limitation of the scaling relationships used here (Equations 2 -- 6) is that they are defined for impacts onto a planar surface.  More studies of experiments and numerical simulations of impacts on curved surfaces are required to define similar scaling laws appropriate for large impacts on a parent body.  Therefore, for very large craters (where the size of the crater approaches the size of the target body) this scaling should be used with caution.  Numerical cratering calculations have confirmed this relationship for craters of diameters up to the size of the planetary radius \citep{Potter:12}.

Another possible outcome of the collision event is the catastrophic disruption of the target.  Catastrophic disruption is defined as a collision in which the largest fragment has a mass of less than half of the original target body mass.  The quantity \qstard\ is defined as the minimum specific impact energy (energy per unit mass of the target) required to disrupt the parent body:

\begin{equation}
Q_D^{\star}=\dfrac{m_{imp}v_{imp}^2}{2m_t}=\dfrac{\rho_{imp}\left(r_{imp}^{\star}\right)^3v_{imp}^2}{2\rho_{t}r_{t}^3}
\end{equation}

Hence, the disruptive impactor radius, \rstari, can be expressed in terms of the impact energy and the target size:

\begin{equation}
r_{imp}^{\star}=r_t\left(\frac{2\rho_t Q_D^{\star}}{\rho_{imp}v_{imp}^2}\right)^{\frac{1}{3}}
\end{equation}

Replacing \rstari\ with \rimp\ in Equation 8 gives the definition of specific impact energy, $Q$:

\begin{equation}
Q=\dfrac{\rho_{imp} v_{imp}^2 }{2\rho_t}\left(\dfrac{r_{imp}}{r_t}\right)^3
\end{equation}

\citet{Benz:99} used numerical modeling to show that \qstard\ can be expressed by the functional form:

\begin{equation}
Q_D^{\star}=Q_0\left(\dfrac{r_t}{0.01\rm{m}}\right)^a+B\rho_t\left(\dfrac{r_t}{0.01\rm{m}}\right)^b
\end{equation}

where $Q_0$, $a$, $b$ and $B$ are material specific constants. The first term on the right-hand side of Equation (10) describes the strength regime, and is only applicable for parent bodies up to approximately $r_t$ = 100 m. Hence, for the parent bodies considered in this work, only the second term on the right-hand side needs to be considered. Several recent studies have attempted to quantify the disruption threshold for a range of scenarios \citep[e.g.][]{Benz:99,Leinhardt:09,Stewart:09}. One such study investigated the effect of porosity on this important parameter: \citet{Jutzi:10} used SPH modeling to simulate planetesimal collisions, and determined material constants for use in Equation (10). For impacts into a porous pumice target at 5 km/s, they found that $B$ = 5.70 erg cm$^3$/g$^2$, and $b$ = 1.22. The explicit porosity is not defined in that work, but the density is approximately half that of the non-porous target material. While pumice is a very different material to basalt or dunite, as this is the only disruption criterion in the literature to incorporate porosity, it is thus the best approximation for use in our model of the early Solar System, when porosity would have been significantly higher than today. For impacts into a non-porous basalt, they found $B$ = 1.50 erg cm$^3$/g$^2$, and $b$ = 1.29. To apply this disruption criterion to a wider range of collision speeds, velocity scaling from \citet{Housen:90} can be applied to the gravity regime term from Equation (10):

\begin{equation}
Q_D^{\star}=B\rho_t\left(\dfrac{r_t}{0.01\rm{m}}\right)^b\left(\dfrac{v_{imp}}{5000 m/s}\right)^{2-b}
\end{equation}

By combining Equations (8) and (11), the minimum disruptive impactor radius can be expressed in terms of the target radius, the impact velocity, and the material specific constants of $b$, $B$, $\rho_t$ and $\rho_i$:

\begin{equation}
r_i^{\star}=\left(2B\dfrac{\rho_r^2}{\rho_{imp}}\left(0.01 \rm{m}\right)^{-b}\left(5000 \rm{m/s}\right)^{b-2}r_t^{3+b}v_{imp}^{-b}\right)^{\frac{1}{3}}
\end{equation}

For each collision in our Monte Carlo model, Equation (12) is used to determine if the collision will catastrophically disrupt the parent body. If disruption does occur, the calculation on that parent body is stopped: No further collisions are modeled.

\subsubsection*{Collision heating}

\begin{table}[t]
\footnotesize
\caption{Peak shock pressures, entropies and specific internal energy increases corresponding to the post-shock temperatures used in this work$^{\rm{a}}$.}
\begin{tabular}{cccc}
\toprule
Post shock  & Specific  & Peak shock & Specific \\
temperature & entropy   & pressure   & internal energy \\
\phantom{.} [K] \phantom{.} & [erg/g/K] & [GPa]      & increase [erg/g] \\
\midrule
350	&	12925	&	1.4	&	$5.29\times10^5$	\\
400	&	14365	&	2.9	&	$1.07\times10^6$	\\
500	&	16810	&	5.8	&	$2.16\times10^6$	\\
600	&	18832	&	8.9	&	$3.27\times10^6$	\\
700	&	20552	&	11.9	&	$4.39\times10^6$	\\
800	&	22048	&	15	&	$5.51\times10^6$	\\
900	&	23371	&	18.1	&	$6.63\times10^6$	\\
1000	&	24557	&	21.3	&	$7.76\times10^6$	\\
1250	&	27070	&	28.9	&	$1.06\times10^7$	\\
1373$^{\rm{b}}$	&	28128	&	32.7	&	$1.20\times10^7$	\\
1500	&	29126	&	36.5	&	$1.34\times10^7$	\\
1750	&	30867	&	43.7	&	$1.62\times10^7$	\\
2000	&	32377	&	50.8	&	$1.90\times10^7$	\\
2053$^{\rm{c}}$	&	32673	&	52.3	&	$1.96\times10^7$	\\
\bottomrule
\end{tabular}
\strut\\
$^{\rm{a}}$The values in this table are derived from the ANEOS equation of state for dunite \citep{Benz:89} and the $\varepsilon-\alpha$ porous compaction model \citep{Wunnemann:06}, using the technique described in \citet{Davison:10a}. These values are appropriate for 20\% porous dunite.\\
$^{\rm{b}}$Solidus\\
$^{\rm{c}}$Liquidus

\end{table}

For each collision that occurs, the amount of heat deposited in the target body can also be estimated. Numerical studies of collisions between planetesimals have quantified the amount of heating done between a range of planetesimal pairs, for a range of collision velocities up to \textsim~10 km/s \citep{Davison:10a}, and found that for an impactor of less than one tenth of the mass of the target, the mass of material shock heated to a given temperature was a constant multiple of the impactor mass (where the constant depends on the impact velocity, target porosity and starting temperature). For impactors greater than one tenth of the mass of the target, the heated mass was limited by the total mass of the target. Here we extend those calculations to consider impact velocities up to 50 km/s (greater than the maximum collision velocity expected from the collisional evolution models discussed above).

We used the iSALE-2D shock physics model \citep{Wunnemann:06}, which is based on the SALE hydrocode \citep{Amsden:80}. To simulate hypervelocity impact processes in solid materials SALE was modified to include an elasto-plastic constitutive model, fragmentation models, various equations of state, and multiple materials \citep{Melosh:92, Ivanov:97}. More recent improvements include a modified strength model \citep{Collins:04} and a porosity compaction model \citep{Wunnemann:06, Collins:11b}. \citet{Davison:10a} further adapted iSALE to allow the simulation of collisions between two planetesimals.  iSALE shock physics calculations were performed for collisions between a 10 km radius impactor into a 100 km radius target body (a mass ratio of 1/1000), for non-porous and 20\% porous planetesimals.  The planetesimals were modeled as spheres, using the ANEOS equation of state tables for dunite \citep{Benz:89}. To determine the mass of material shock heated to several different temperatures, Lagrangian tracer particles were placed in the computational mesh.  The mass of all particles that experienced a threshold peak pressure required to reach the desired final, post-shock temperature was determined in post-processing \citep{Pierazzo:97, Pierazzo:00, Ivanov:02, Davison:10a}. Figure 3 shows the mass of material heated to several temperatures for 20\% porosity, for the full range of velocities considered in this study.  To determine the heated mass in the Monte Carlo model, a linear interpolation between the two closest velocities is used. If the heated mass interpolated from the iSALE models exceeds the mass of the parent body, the heated mass recorded from this collision is limited to the parent body mass. The increase in specific internal energy for each temperature increase is also listed in Table 1. For each impact, therefore, the total amount of energy that is used for heating the parent body, \qheat, can be approximated by:

\begin{multline}
Q_{heat}=\Delta Q_{2053}M\left(>T_{2053}\right)+\\ 
\sum_{j=1}^{n-1}\Delta Q_j\left(M\left(>T_j\right)-M\left(>T_{j+1}\right)\right)
\end{multline}

where $\Delta Q_j$ is the specific internal energy increase associated with shock heating the material to the post-shock temperature $T_j$ (listed in Table 1), $M\left(>T_j\right)$ is the mass heated to the given temperature in the collision (Figure 3). \qheat\ can be converted to a specific internal energy increase for the parent body by dividing by the parent body mass; however it should be noted that \qheat\ will be non-uniformly distributed within the parent body Ñ energy will be localized to the impact sites.

\begin{figure}[!t]
\includegraphics[width=0.9\linewidth]{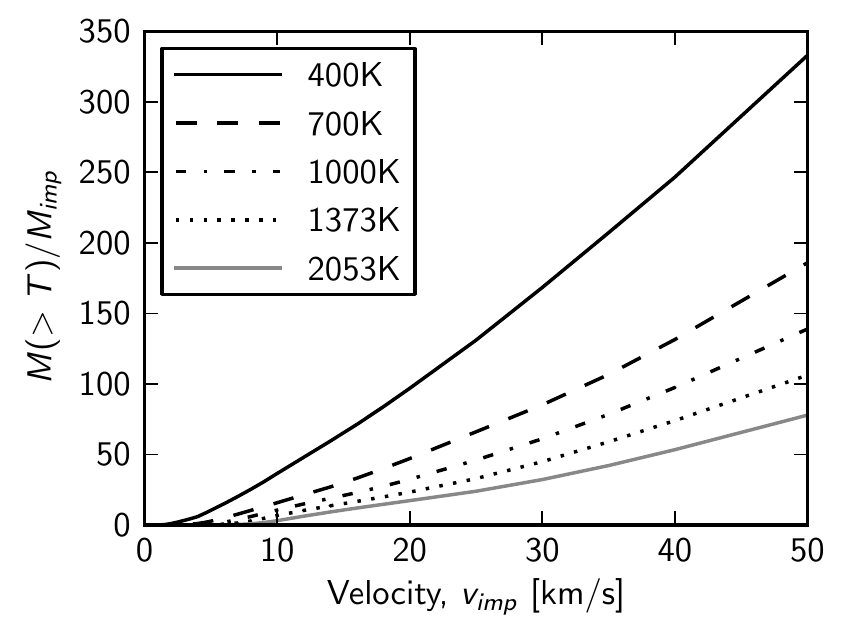}
\caption{Mass of material heated in a collision between two 20\% porous dunite planetesimals, where the impactor has one tenth the radius of the target (one thousandth of the mass), for a range of impact velocities. The heated mass is normalised to the mass of the impactor. Both planetesimals had an initial temperature of 300 K. $T$ = 1373 K and 2053 K are the inferred dunite solidus and liquidus, respectively.}
\end{figure}

The techniques used here to determine the final temperature of the material are dependent on the accuracy of the equation of state for converting peak shock pressures to post-shock temperatures. As the ANEOS equation of state does not account for latent heat, ANEOS over-estimates temperatures in excess of the melt temperature. To account for this, the peak shock pressures and entropy that correspond to the post-shock temperatures that are used in this work are presented in Table 1. As ANEOS equations of state are improved in the future (for example by defining different thermodynamic constants to be used for different high-pressure phases, and by treating each phase as a separate material in ANEOS), these shock pressures and entropies can be used to amend the temperatures quoted in the remainder of this work. However, as shock melting is minimal in most cases presented here, the latent heat effects introduce only small uncertainties to our results.

\subsubsection*{Timestep}
After these calculations have taken place for all impactor size bins, the timestep is advanced and the process is repeated, unless a disruptive impact has occurred, in which case the calculation for this parent body ceases, and the calculation of the collisional history of the next parent body starts.  The size of the timestep is chosen so that the probability of a collision by an impactor in the smallest size bin, $P_c\left(r_{min},t\right)$ (which is the most likely collision to occur), is less than unity.  To explore the sensitivity of the results to the size of the timestep, simulations were run with different timestep values, so that $P_c\left(r_{min},t\right)$ = 0.1, 0.25, 0.5 and 1.  For $P_c\left(r_{min},t\right)$ = 1, the total number of collisions experienced by the parent body was less than for $P_c\left(r_{min},t\right)$ ² 0.5, but did not change for $P_c\left(r_{min},t\right)$ = 0.1 -- 0.5.  Therefore a timestep limit of $P_c\left(r_{min},t\right)$= 0.5 was used for all simulations presented here.  This approach is an approximation of a Poisson distribution.

\subsection*{Assumptions and limitations}
\subsubsection*{Time period}
The model described above focuses on the early impact evolution of the Solar System.  Only the first 100 Myr of Solar System evolution are accounted for, as this is the time period in which the greatest number of impacts occurred.  By 100 Myr, the intrinsic collision probability and the total mass of planetesimals are significantly lower than at the start of the calculation.  As the results below show, the frequency of collisions at t $\approx$ 100 Myr is much lower than in the first \textsim~20 Myr. Collisions after this time are not accounted for, but in this study we are focused on the earliest effects of collisions when impacts were most frequent. For future studies interested in a later period of Solar System history, the collisional evolution models could be run for a longer period of time, and the same Monte Carlo methods applied to those results.

\subsubsection*{The parent body}
To fully quantify the amount of heating possible from multiple impacts on the surface of a meteorite parent body, no other heat sources (e.g. $^{26}$Al decay) were considered in this study.  The parent body was therefore assigned a constant initial temperature of 300 K, which is the reference temperature of dunite from the ANEOS equation of state \citep{Benz:89, Thompson:72}, used in hydrocode models to determine heating for specific impact scenarios \citep{Davison:10a}. Such a temperature may be appropriate for a planetesimal inside the snow line in the early Solar System, and therefore this assumption does not affect many of the results of this paper.  However, crater size and impact heating calculations are dependent on the initial temperature of the material.  Thus, to extend this study to a wider region of the Solar System, future studies investigating this parameter in more detail are required.

The conditions for solid materials in the early Solar System are somewhat uncertain. It is likely that the earliest solid bodies contained a significant fraction of pore space \citep{Bland:11, Blum:03, Dominik:97, Love:93, Wurm:01, Wurm:04}. The gentle manner in which dust grains must collide during planet formation likely leads to bodies with high porosities \citep{Teiser:09, Blum:08}.  Even growth to planetesimal formation must occur at very low velocities to ensure sticking \citep[e.g.][]{Weidenschilling:80, Weidenschilling:84, Weidenschilling:97} or to allow aggregates to become gravitationally bound \citep{Cuzzi:01, Cuzzi:08, Johansen:07}.  Indeed, it is thought that the porous nature of early planetesimals is seen in the fabric of the Allende meteorite, with initial porosities reaching 60 -- 70\% \citep{Bland:11}.

Once the planetesimal formed, porosity would diminish:  The rate of densification is a strong function of internal pressure and temperature \citep{Schwenn:78, Yomogida:84, Henke:12}.  To simplify the calculations in this work, it is assumed that all parent bodies and impacting planetesimals modeled have a constant porosity that does not change with time.  It is possible to estimate the size of parent body for which porosity would be crushed out due to the internal pressure at the center.  Assuming the bulk density, $\rho_t$ remains constant throughout the parent body, then the pressure, $P$, at a distance $r$ from the center of a body with radius \rt\ can be expressed as \citep{Turcotte:02}:

\begin{equation}
P=\dfrac{2}{3}\pi\rho_t^2 G\left(r_t^2-r^2\right)
\end{equation}

For a 20\% porous dunite parent body with a radius of 250 km, the internal pressure at the center would be approximately 61 MPa, and for a body with a radius of 100 km, the pressure would be approximately 10 MPa. \citet{Nakamura:09} performed crush experiments on a range of porous materials to determine the pressure at which pore space begins to collapse. For an initially 50\% porous gypsum sample, crushing began at 20--30 MPa, and at 50 MPa the porosity had been reduced to 40\%. An initially 40\% porous sample experienced little reduction in pore space, even after 100 MPa of pressure was applied. Indeed, \citet{Menendez:96} show that static pressures of 500 MPa to 1 GPa are required to lithify sandstones on Earth. Therefore, the reduction of pore space in a cold parent body due to internal pressure alone is not likely to be significant. Densification would thus be minimal in bodies of the size of interest, unless temperatures were elevated to $>$ 700 -- 800 K, at which temperatures sintering by creep diffusion would allow for densification on timescales of \textsim~10$^6$ years. At lower temperatures, primordial porosity would be preserved. Even with radiogenic heating, porous layers will be preserved in the outer regions of planetesimals (which are most affected by impact) as a result of the low hydrostatic pressures and temperatures expected in that region \citep{Henke:12}. Studies of asteroid densities show that most asteroids contain both macro- and microporosity \citep[e.g.][]{Britt:02, Britt:03, Consolmagno:04, Consolmagno:08}, suggesting that at least some of the initial (micro) porosity of a parent body would be maintained through time, along with the introduction of new (macro) porosity due to fracturing and breakup during impact events. Each impact on a parent body could also crush out some pore space. This compaction would be constrained to the locality of the impact site \citep{Wunnemann:06, Davison:12}. Hence, in the simulations presented here it is assumed that each impact occurs on a fresh, uncompacted target surface. Until we have better data on the evolution of porosity through time, this assumption is a necessary simplification in the model.

\subsubsection*{Mass of the parent body}
It is assumed that ejection velocities are suppressed during impact into a porous target relative to those expected during impact into a non porous target \citep{Collins:11c}, and therefore that all material heated during a planetesimal collision is retained on the surface of the parent body (unless the parent body is catastrophically disrupted; see below).  As a consequence of this assumption, the mass of the parent body remains the same throughout the simulation (the mass lost to ejection is neglected).  While this is an over-simplification, it is not expected to affect the mass of heated material on the parent body, as it is likely that most heated mass is not ejected during the collision \citep{Collins:11c}.  In addition, the mass added from impactors is not accounted for in the simulation.  As most impactors are small in comparison to the parent body (and those that are large are more likely to lead to a disruption), this assumption is appropriate for a first-order approximation.  In the simulations discussed below, the average total volume of impactors on a parent body ranges from 1 -- 20\%, depending on the parent body size and collisional evolution model.  Future work incorporating collisional outcome calculations \citep[e.g.][]{Davison:10a, Leinhardt:12} into the Monte Carlo simulation will address this limitation.

\subsubsection*{Impact angle}
For each collision that takes place, no impact angle is assigned. For calculations of the disruption threshold, an impact angle of 45¡ is assumed, as it is the most probable impact angle \citep[see above, and][]{Jutzi:10}. For calculations of crater dimensions and impact heating, the impact angle is assumed to be vertical. This limitation is because much of the work defining the scaling laws was performed using two- dimensional numerical simulations, which employed axial symmetry. Vertical incidence impacts produce craters that are only \textsim~10\% larger than impacts at 45\degr \citep{Gault:78, Elbeshausen:09, Davison:11}. Preliminary modeling work has shown that the mass of material heated in an impact at 45\degr may be approximately 20--30\% less than in vertical incidence impacts \citep{Davison:12}. As the use of three-dimensional simulations becomes more common (as computational speeds improve), the effect of impact angle on crater size and impact heating can be quantified, and subsequently used to overcome this limitation in this work. This assumption only affects our estimates of the mass heated in the planetesimal and does not affect the size of the impactors or impact velocities that will be experienced by a given body.

\msection{RESULTS}
There exists a large parameter space to which this approach can be applied. In the following section, the results of the Monte Carlo model described above are presented for the collisional history of a 100 km radius target body for the CJS model of collisional evolution, with $f$ = 100. A radius of 100 km is representative of the expected size of chondrite parent bodies \citep[e.g.][]{Harrison:10a}. In discussing the thermal consequences and disruption threshold of the impacts, we assume a uniform porosity of 20\%, a value that is less than the expected initial porosity of fresh planetesimals and close to the typical porosity seen in current chondrites. While porosity likely evolves in complicated ways due to compaction from heating and impacts as well as creation of pore space by impact fracturing and shear deformation, we ignore this evolution for simplicity. In subsequent sections, we investigate the sensitivity of a parent bodyÕs collisional history to the disk collisional evolution model (CJS or EJS), the size of the parent body (\rt = 50, 100 and 250 km) and the porosity (non-porous and 20\%) of the parent body.

\begin{figure}[!htp]
\includegraphics[width=0.9\linewidth]{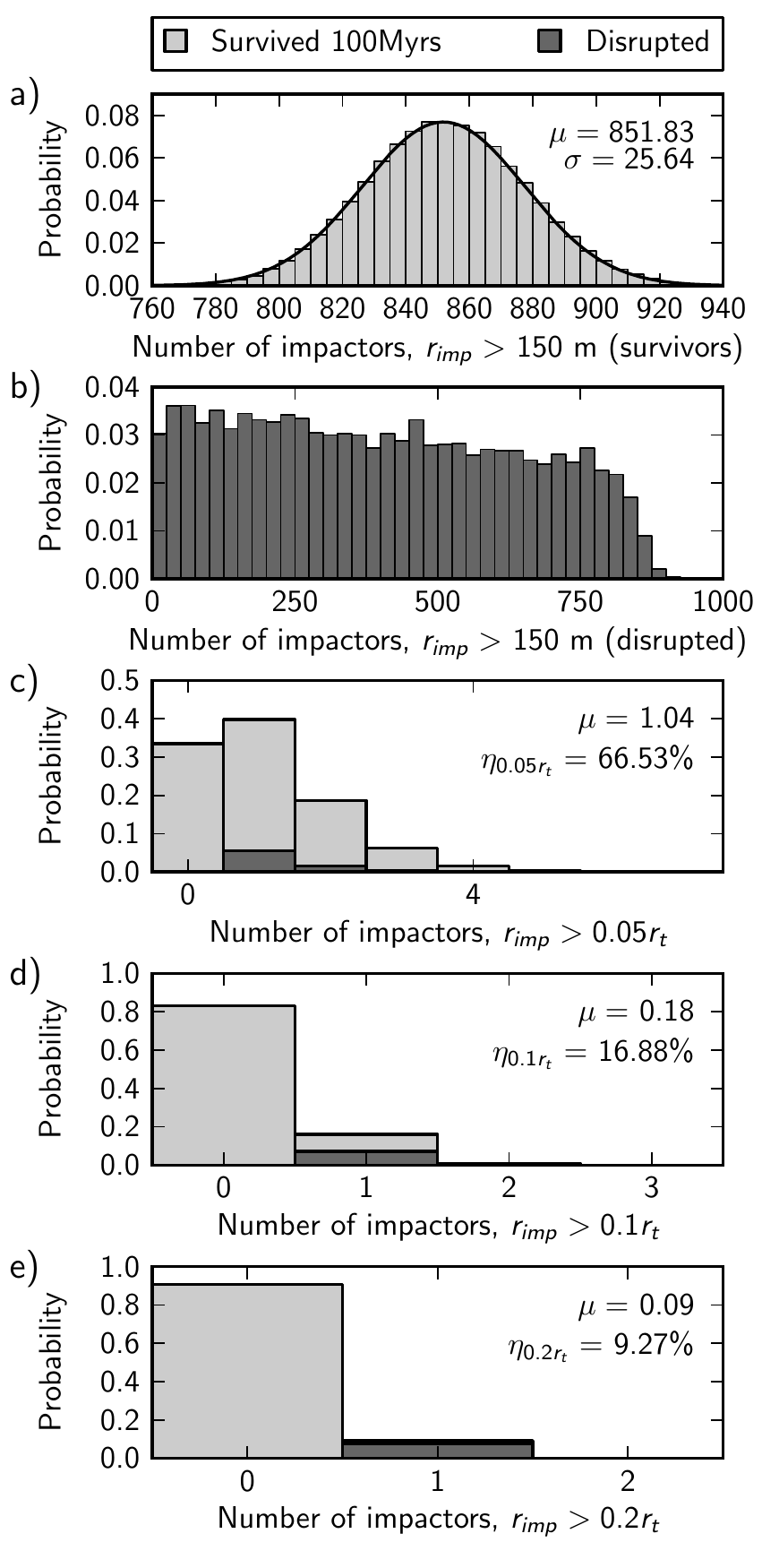}
\caption{Statistics of the number of impacts expected on a 100 km radius, 20\% porous parent body, from the CJS model. a) The probability that a parent body that survives to 100 Myr will receive a given number of impacts from impactors with radii $>$ 150 m; $\mu$ and $\sigma$ are the mean and standard deviation. b) The probability that a parent body that is disrupted before 100 Myr will receive a given number of impacts from impactors with radii $>$ 150 m. c--e) The probability that a parent body will receive a given number of impacts from impacts larger than a certain size. $\eta$ is the probability that a parent body will experience at least one impact of this size or larger.}
\end{figure}

\subsection*{CJS model, \rt = 100 km, $\phi$ = 20\%}
\subsubsection*{Overview}
Figure 4 shows statistics of the number of impacts experienced by a 20\% porous, 100 km radius parent body. On average, a parent body that survives 100 Myr without being catastrophically disrupted sustains \textsim~851 $\pm$ 26 (1-$\sigma$ deviations) impacts from bodies with a radius greater than 150 m (Figure 4a). During this same time period, 7.6\% of parent bodies experience a disruptive collision: i.e. they experience a collision for which Q $>$ \qstard. Of those that are disrupted, the number of collisions with \rimp $>$ 150 m prior to disruption varies from 1 to \textsim~900, with an approximately equal chance of receiving any number of collisions within that range (Figure 4b). Figures 4c--e show the number of collisions sustained by the parent body for impactors larger than 0.05 \rt, 0.1 \rt\ and 0.2 \rt, respectively. On average, a parent body receives \textsim~1.04 impacts with \rimp\ $>$ 0.05 \rt. The probability that a parent body will experience at least one impact of this size or larger (defined here as \etan{0.05}) is 66\%. Approximately one in five parent bodies experience an
impact with \rimp\ $>$ 0.1 \rt (\etan{0.1} = 17\%) and of that number, approximately half are disrupted. The mean number of impacts for which \rimp\ $>$ 0.2 \rt is 0.09, and \etan{0.2} = 9\%.

Of those that experience at least one impact in this size range, around 4 out of 5 were disrupted---hence, the one in five that was not disrupted experienced a collision with a large impactor at sufficiently low velocity to allow survival. A summary of these statistics is presented in Table 2, along with statistics from several other simulations.

\begin{table*}[!t]\footnotesize
\caption{Statistics for disruptive events in Monte Carlo simulations for a range of parent body sizes and disk models, after $10^5$ parent bodies had been modeled for each case.}
\begin{center}
\begin{tabular}{cccrrrrc}
\toprule
\multirow{3}{*}{Model} & \multirow{3}{*}{f $^{\rm{a}}$} & & \multicolumn{4}{c}{\Nimp $>$ \rimp $^{\rm{c}}$} & \\
\cmidrule(r){4-7}
& & \rt $^{\rm{b}}$ & 150m & \multirow{2}{*}{0.05\rt} & \multirow{2}{*}{0.1\rt} & \multirow{2}{*}{0.2\rt} & $N_{disrupt}$ $^{\rm{d}}$ \\
& & [km] & (survivors) & & & & \% \\
\toprule
\multirow{2}{*}{CJS} &\multirow{2}{*}{100} &\multirow{2}{*}{50} & $\mu$ = 214.25 & $\mu$ = 1.56 & $\mu$ = 0.32 & $\mu$ = 0.07 & \multirow{2}{*}{8.2}\\ 
&&&	$\sigma$ = 13.96 & $\eta$ = 80.5\% $^{\rm{e}}$ & $\eta$= 28.2\% & $\eta$ = 7.3\%\\
\midrule	
\multirow{2}{*}{CJS} &\multirow{2}{*}{100} &\multirow{2}{*}{100} & $\mu$ = 851.83 & $\mu$ =  1.04 & $\mu$ = 0.18 & $\mu$ = 0.09 & \multirow{2}{*}{7.6}\\ 
&&&	$\sigma$ = 25.64 & $\eta$ = 66.5\%  & $\eta$= 16.9\% & $\eta$ = 9.3\%\\
\midrule
\multirow{2}{*}{CJS} &\multirow{2}{*}{100} &\multirow{2}{*}{250} & $\mu$ = 5303.38 & $\mu$ = 0.81 & $\mu$ = 0.36 & $\mu$ = 0.17 & \multirow{2}{*}{2.6}\\ 
&&&	$\sigma$ = 61.84 & $\eta$ = 55.9\%  & $\eta$= 30.6\% & $\eta$ = 15.6\%\\
\toprule
\multirow{2}{*}{EJS} &\multirow{2}{*}{95} &\multirow{2}{*}{50} & $\mu$ = 77.28 & $\mu$ = 0.71 & $\mu$ = 0.15 & $\mu$ = 0.03 & \multirow{2}{*}{3.7}\\ 
&&&	$\sigma$ = 8.55 & $\eta$ = 51.8\%  & $\eta$= 14.0\% & $\eta$ = 3.8\%\\
\midrule
\multirow{2}{*}{EJS} &\multirow{2}{*}{95} &\multirow{2}{*}{100} & $\mu$ = 307.35 & $\mu$ = 0.48 & $\mu$ = 0.07 & $\mu$ = 0.03 & \multirow{2}{*}{2.9}\\ 
&&&	$\sigma$ = 15.86 & $\eta$ = 39.1\%  & $\eta$= 6.5\% & $\eta$ = 3.3\%\\
\midrule
\multirow{2}{*}{EJS} &\multirow{2}{*}{95} &\multirow{2}{*}{250} & $\mu$ = 1912.58 & $\mu$ = 0.30 & $\mu$ = 0.13 & $\mu$ = 0.06 & \multirow{2}{*}{1.3}\\ 
&&&	$\sigma$ = 37.50 & $\eta$ = 26.0\%  & $\eta$= 11.9\% & $\eta$ = 6.1\%\\
\bottomrule
\\
\multicolumn{8}{p{12cm}}{$^{\rm{a}}$ $f$ is the ratio of the total initial mass of the planetesimal population to the final mass of planetesimals in the asteroid belt.}\\
\multicolumn{8}{p{12cm}}{$^{\rm{b}}$ \rt\ is the radius of the parent body.}\\
\multicolumn{8}{p{12cm}}{$^{\rm{c}}$ \rimp\ is the impactor radius.}\\
\multicolumn{8}{p{12cm}}{$^{\rm{d}}$ $N_{disrupt}$ is the percentage of parent bodies catastrophically disrupted within 100 Myr.}\\
\multicolumn{8}{p{12cm}}{$^{\rm{e}}$ $\eta$ is defined as the probability that a parent body will experience at least one impact greater than or equal to \rimp. These figures are for parent bodies with 20\% porosity.}
\end{tabular}
\end{center}
\end{table*}

\subsubsection*{Crater sizes, excavation depth and impact energy from individual collisions}
In addition to impactor size, the energy (velocity) of the impact is critical for determining the extent to which materials are processed during a collision. For each of the impacts that occurred, we also determined the energy as a fraction of the disruptive energy (i.e. \qdqstard ). Figure 5a shows the mean cumulative number of impacts on a parent body as a function of \qdqstard. On average, each parent body will see one impact with a \qdqstard\ of at least 0.005. These impacts may be from relatively large impactors moving at low velocity or small impactors moving at very high velocities---this figure does not differentiate between them.

\begin{figure}[!t]
\includegraphics[width=0.9\linewidth]{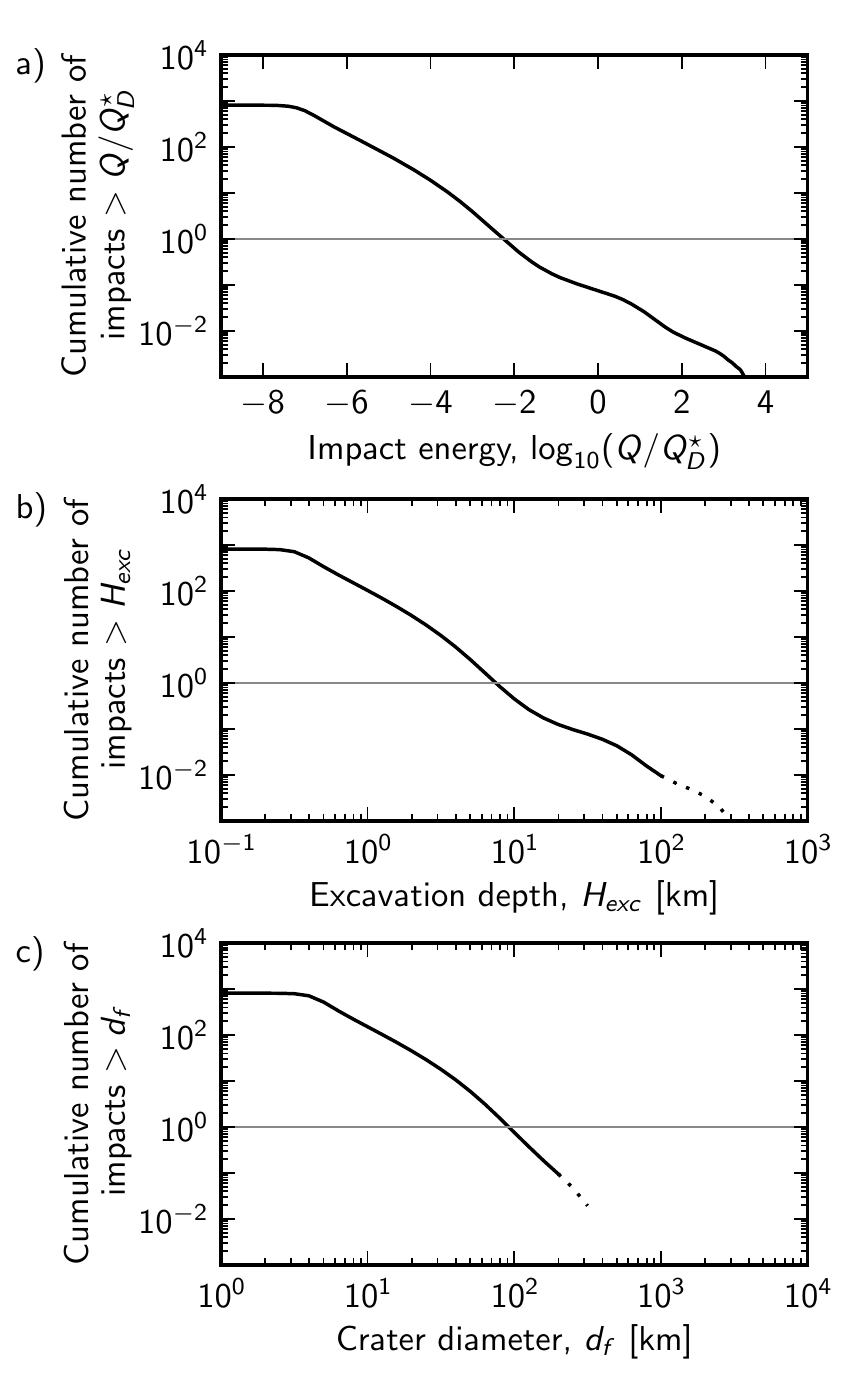}
\caption{Plots showing the average number of collisions that each parent body experiences as a function of the scale of the impact. (a) Cumulative number of impacts with an impact energy greater than \qdqstard; (b) Cumulative number of impacts that excavate material from greater than $H_{exc}$ depth; (c) Cumulative number of impacts that form a final crater larger than $d_f$. Dotted lines in (b) and (c) represent events for which the scaling laws used are not sufficient to describe the outcome of the impact: i.e. an excavation depth greater than the radius of the parent body or a crater diameter larger than the parent body diameter are not physical; these events likely represent either disruptive or sub-catastrophic events, which may not form a traditional impact crater.}
\end{figure}

For each impact that occurs on the parent body, the depth that material is excavated from and the size of the impact crater formed can be determined from scaling laws (Equations 2 -- 6). Figure 5b shows the number of impacts on a parent body as a function of the excavation depth: Each parent body has a 50\% chance of experiencing at least one impact that excavates material from at least 10 km depth, while \textsim~100 impacts will excavate material from a depth of at least 1 km. Figure 5c shows the number of impacts on a parent body as a function of the final crater diameter. The transition from simple to complex craters on a 100 km radius, 20\% porous parent body occurs at \Df $\approx$ 320 km: i.e. larger than the parent body itself; in other words, all craters that form on this parent body, and do not catastrophically disrupt it, will be simple craters. On average, a parent body will experience one impact that forms a crater at least 94 km in diameter. Approximately 150 events per parent body will form craters 10 km in diameter, or larger. On those parent bodies that are not disrupted during the initial 100 Myr, on average the total area of all craters is \textsim~0.8 times the surface area of the parent body (assuming the craters are circular, with a final diameter calculated from Equations 2 -- 5). A representative example of the surface from one surviving parent body is shown in Figure 6. In this figure, the craters were placed randomly on a square surface with an area equal to the parent body surface area. Figure 6 shows that while some craters will overlap one another, a large number of the craters will form at least partially on a fresh, uncompacted surface, and that only a small portion of the surface will be left unaffected by collisions. Thus, the assumption of the model that all craters form on a porous surface is reasonable. This calculation assumes that each crater is formed by vertical impact, and thus should be viewed as an upper limit. Experiments and three dimensional modeling suggest that at an impact of 45\degr, the crater area may be approximately 20\% smaller in area than a normal incidence crater \citep{Burchell:98, Davison:11}. Figure 5 shows that while not all of the \textsim~850 events that are predicted to occur from the model will have a global effect on the parent body, there are likely to be hundreds of events that will influence the outer few km of the parent body: That is, the region most likely to be the source region for low petrologic type chondrites \cite[e.g.][]{Harrison:10a}. Only a small fraction of the surface area is likely to escape without being directly processed by impacts.

\begin{figure}[!t]
\includegraphics[width=0.9\linewidth]{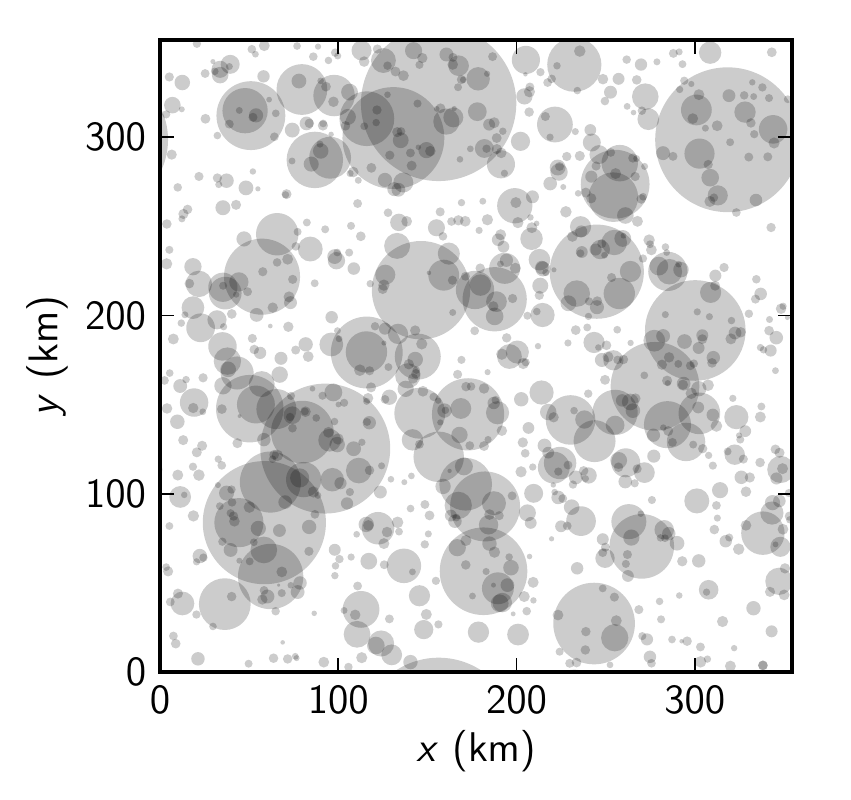}
\caption{The craters from one 100 km radius, 20\% porous parent body from the CJS simulation, randomly placed on a square with an area equal to the surface area of the parent body, to show the coverage of craters on an average surviving parent body. While some craters overlap, most form at least partially on a fresh surface. Darker shading represents overlapping craters.}
\end{figure}

\subsubsection*{Cumulative impact energy}
The total amount of impact energy deposited on a parent body can be determined by summing together the impact energy, $Q$, for each impact. For 99\% of the parent bodies that survive to 100 Myr in this simulation, the total impact energy, \Qtot, is within the range $4.5\times10^7$ -- $6.3\times10^9$ erg/g, with a geometric mean \Qtot\ of $2.3\times10^8$ erg/g. For parent bodies that are disrupted, \Qtot\ is several orders of magnitude larger, mainly due to the energy of the final disruptive impact: 99\% of disrupted bodies have a \Qtot\ in the range $2.8\times10^9$ -- $5.1\times10^{13}$ erg/g, and the geometric mean \Qtot\ is $6.0\times10^{10}$ erg/g. For reference, \qstard\ from \citet{Jutzi:10} for an impact at 5 km/s into a porous 100 km parent body is $\sim 2\times10^9$ erg/g. Figure 7a shows a histogram of \Qtot\ for all parent bodies modeled. A fraction of this total impact energy is converted to heat; this is quantified below, in the section `\emph{Impact heating}'.

\begin{figure}[!t]
\includegraphics[width=0.9\linewidth]{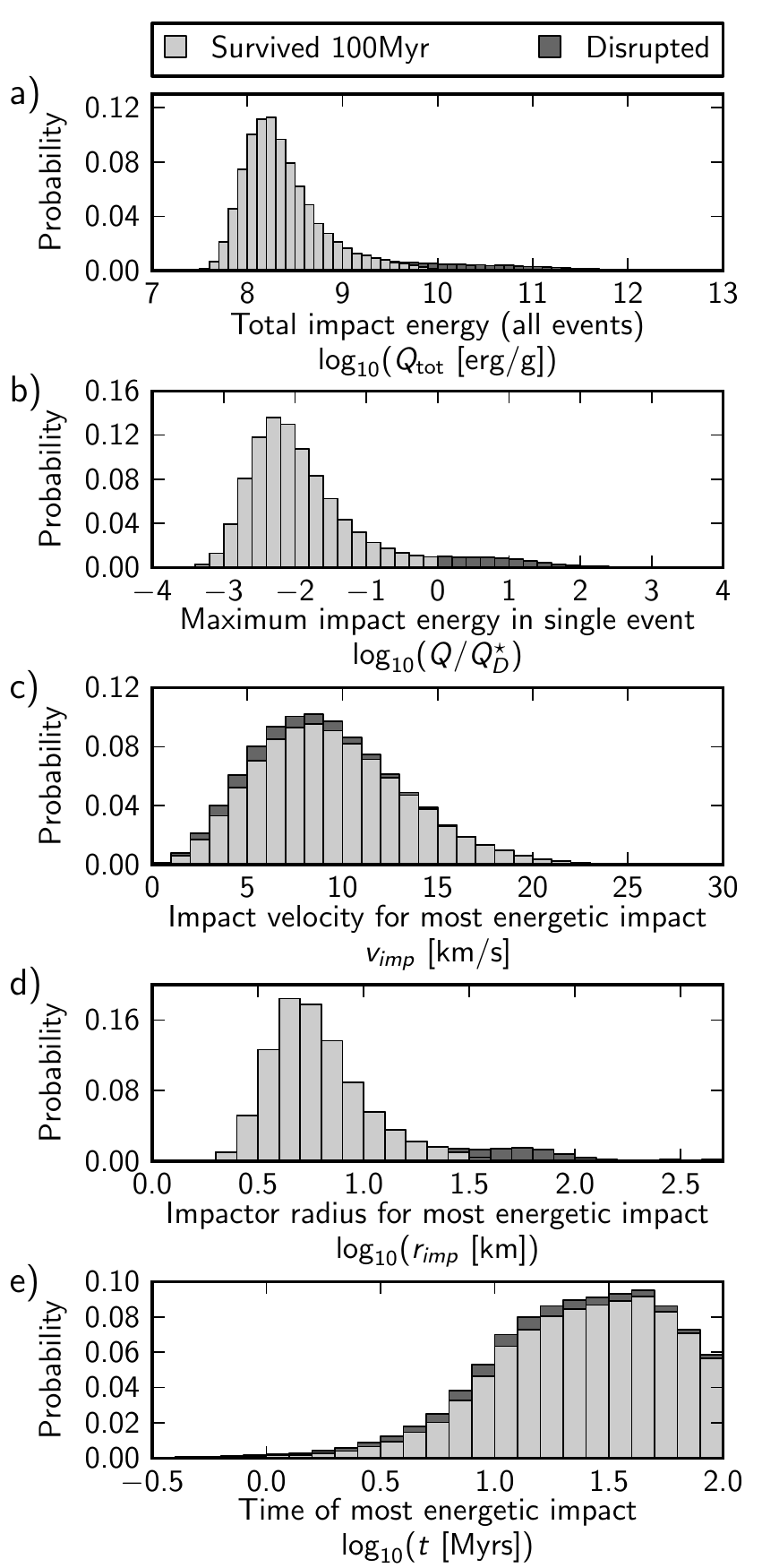}
\caption{Statistics for the amount of impact energy delivered to a 100 km radius, 20\% porous parent body from the CJS simulation. The `most energetic impact' is defined as the impact with the maximum \qdqstard on each parent body. \Qtot\ is the total impact energy received on the parent body in the first 100 Myr.}
\end{figure}

\subsubsection*{Most energetic impact}
As \qstard\ depends on the impact velocity and the impactor size, the impact with the largest $Q$ may not be the impact closest to the disruption threshold. Indeed, in approximately 1 out of every 8 parent bodies, the most energetic impact is not the most disruptive. Therefore, the most energetic impact sustained on a parent body is defined here as the impact with the largest \qdqstard. Figure 7b shows the distribution of the most energetic impact for all parent bodies modeled. For those parent bodies that are not disrupted, \qdqstard\ varies between \textsim~10$^{-3}$ to 1, and the geometric mean is 0.01; i.e. the average impact energy of the most energetic impact in parent bodies that are not disrupted is approximately 1\% of that needed to disrupt the body. For reference, \qdqstard\ = 0.01 is approximately equivalent to a 7.5 km radius projectile impacting the 100 km radius parent body at 5 km/s. The collateral effects of this type of impact are discussed in more detail in \citet{Davison:12}; such impacts can produce peak temperatures and cooling rates consistent with what is seen in type 3, 4, 5, and 6 H-chondrite meteorites. Figure 7c-d shows the impact velocity and radius for the most energetic impact. A typical impact velocity in the most energetic impact is \textsim~7--10 km/s, but a range from \textsim~1 km/s up to 25 km/s is possible. For bodies that are disrupted, the impactor radius is typically in the range 30 -- 100 km. The minimum disruptive impactor radius in the simulations we performed is \textsim~22.2 km (for which the impact velocity was 17.8 km/s), and 99\% of all disruptors are $>$ 28 km in radius. For those bodies that are not disrupted in the first 100 Myr, the average radius of the most energetic impactor is \textsim~7 km. For nearly 1\% of surviving parent bodies, the most energetic impactor was $>$ 30 km in radius; these impactors have low velocities so the impact energy is below \qstard.

\subsubsection*{Time of impact}
Figure 8 shows the average number of collisions on a potential meteorite parent body (impactors larger than \rimp\ = 150 m) during each 1 Myr period of the first 100 Myr. \textsim~19\% of the collisions that a parent body experienced were during the first 5 Myr, \textsim~39\% during the first 10 Myr, and \textsim~64\% during the first 20 Myr. 75\% of impacts occur within the first 29 Myr. Within a factor of \textsim~2, the number of impacts on a body in the first 100 Myr is expected to be approximately equal to the number of impacts in the remaining 4.4 Gyr.

\begin{figure}[!t]
\includegraphics[width=0.9\linewidth]{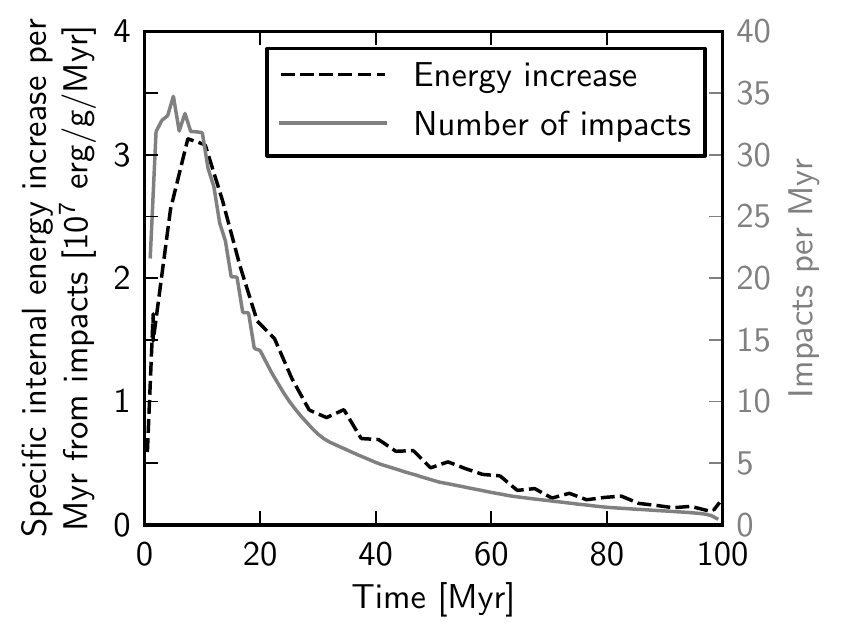}
\caption{The average number of impacts by impactors with radius $>$ 150 m that occurred in each 1 Myr period, for the first 100 Myr (grey solid line), and the specific internal energy increase that occurred due to those impacts (black dashed line). These results are for a 100 km radius, 20\% porous parent body from the CJS simulation. Averages were taken after 100000 parent bodies were modeled. The most important time for impacts was the first 10--20 Myr.}
\end{figure}

Figure 7e shows the time that the most energetic impact occurs for the modeled parent bodies. Approximately 21\% of disruptive collisions occur in the first 5 Myr, 43\% occur in the first 10 Myr, and 68\% occur in the first 20 Myr. Of those parent bodies that are not disrupted in the first 20 Myr, 97\% will survive to the end of the 100 Myr period simulated by this model (for this study, `surviving' is defined as not being disrupted by 100 Myr).

The average time of the most energetic collision in survivors occurs later than the disruptive collisions: Only 5\% of these collisions occur in the first 5 Myr, 16\% occur in the first 10 Myr, and 39\% occur in the first 20 Myr. Some of the most energetic collisions experienced on the parent body therefore continue later than the disruptive collisions, although after \textsim~30--40 Myr, their frequency decreases (only one third of the most disruptive collisions occur after 40 Myr). The difference in the timing of disruptive collisions and most energetic collisions on surviving bodies is because: (1) if a disruptive collision occurs, the simulation is stopped, and no more collisions can occur on that body, which would therefore favor earlier collisions, and (2) while the average collision velocity continues to increase until around 30 Myr, the number of large impactors, capable of causing a disruption, decreases rapidly after \textsim~10 Myr.

\subsubsection*{Impact heating}
Figure 9 shows the probability that a given fraction of a parent body will be heated to a given temperature (assuming the impact occurs onto a fresh surface), for those parent bodies that survived 100 Myr without a disruptive collision (Figure 9(a)), and those that were disrupted (Figures 9b--c). These figures can be interpreted in two ways: First, if one is interested in the outcome associated with a given probability, the figure can be read horizontally, to establish the mass fraction heated to a given temperature (this is the maximum mass fraction, as oblique impacts are not considered). Second, to infer the probability of a particular outcome (e.g. 50\% of the mass being heated to a given temperature), the figure can be read vertically. For example, of the surviving parent bodies, there is a 20\% chance that a parent body will see around one thirtieth of the mass heated to 400 K, and a 30\% chance that one hundredth of the mass will be heated to the solidus. For those parent bodies that are disrupted, however, there is an 80\% chance that the entire parent body will be heated to 400 K, and a 50\% chance that one fifth of the mass of the parent body reaches the solidus. Comparing Figures 9b and 9c, it is clear that the large impact that causes the parent body to disrupt is also by far the most important impact in terms of heating the target materials. Further modeling is required to determine how the heated material will be distributed amongst the collisional fragments in disruptive events.

\begin{figure}[!t]
\includegraphics[width=0.9\linewidth]{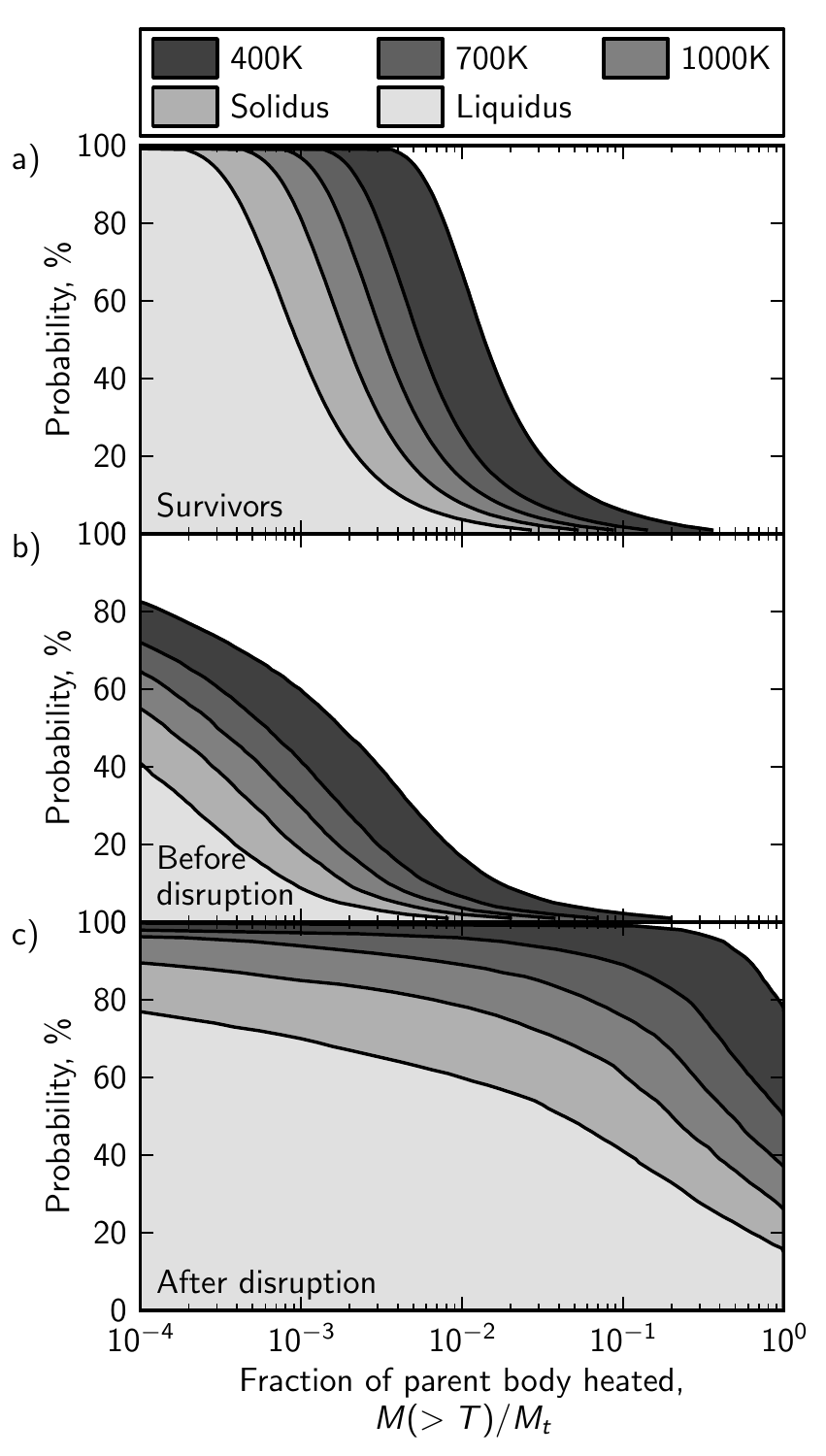}
\caption{The probability that a given fraction of a parent body will be heated by impacts to a range of temperatures, for (a) those bodies that survive 100 Myr without being catastrophically disrupted, (b) those bodies that are disrupted, immediately before the disruptive impact occurs, and (c) those bodies that are disrupted, after the disruptive impact.}
\end{figure}

By using Equation (13), the specific internal energy increase due to collisional heating \Qheat\ can be determined for each parent body. For those parent bodies that survive, the mean specific internal energy increase due to all the collisions that occur in the first 100 Myr is $1.6\times10^8 \pm 3.3\times10^8$ erg/g. This represents \textsim~37.4 $\pm$ 3.6\% of the predicted total impact energy \Qtot\ received by surviving bodies and produces a globally averaged temperature increase of \textsim~15--35 K. This is consistent with earlier findings that non-disruptive collisions are unlikely to cause global heating on a meteorite parent body \citep{Keil:97}. For disrupted parent bodies, much more impact energy is converted to heat: The total combined specific internal energy increase in the target body is $9.6\times10^9 \pm 6.7\times10^9$ erg/g (corresponding to global average temperature increases of \textsim~600--1000 K). However, this represents a smaller percentage of the total impact energy received on those bodies (17.7 $\pm$ 11.8\%), because much of that impact energy is used mechanically to disrupt the parent body. As all the impacts used in this analysis are normal incidence impacts, the estimate of the amount of impact energy converted to heat should be considered an upper limit: Oblique impacts would likely convert less impact energy into heat.

Figure 8 also shows this increase in specific internal energy as a function of time. This reiterates that the initial \textsim~10--20 Myr are the most important in terms of collisional processing and the delivery of heat to the parent body, with the peak increase of $> 3\times10^7$ erg/g/Myr occurring at \textsim~11 Myr. However, this figure also shows that even after 50 Myr, at which time heating from short-lived radionuclides would be negligible, collisions could still deliver $\sim 5\times10^6$ erg/g/Myr. Moreover, as most impact energy is deposited in the outer layers of the parent body (e.g. of the \textsim~850 expected impacts on a parent body, on average less than one impact will excavate material from more than 10 km depth), the specific internal energy increase in that material may be significantly higher. As a first-order approximation, if it is assumed all the specific internal energy is deposited in the outer 10 km of the parent body, then the energies presented above and in Figure 8 could increase by a factor of \textsim~4. This shows that while global heating by impacts is unlikely, heating on the local scale during the period of peak impact activity is possible.

\subsection*{Influence of parent body size}
In addition to the simulation described above, similar models with the CJS collisional evolution model were run with different parent body sizes: \rt\ = 50 km and \rt\ = 250 km.

\subsubsection*{Number of impacts}
Table 2 shows statistics for the number of collisions and the probability of catastrophically disrupting the parent body. A parent body of 50 km radius experiences a factor of \textsim~4 fewer collisions than the 100 km parent body, and a 250 km parent body experiences \textsim~6 times more collisions than a 100 km body -- these factors correspond to the change in surface area. Smaller parent bodies are more likely to be disrupted, despite the lower frequency of collisions: 8.2\% of 50 km radius bodies are disrupted, 7.6\% of 100 km bodies, and 2.6\% of 250 km bodies, due to the \qstard\ dependence on parent body size---the value of \qstard\ increases with target size as we are in the gravity regime.

The number of impactors one twentieth of the radius of the parent body decreases with parent body size: On average, there are \textsim~1.6 impacts of this size on a 50 km parent body, compared to \textsim~1.0 on a 100 km body and \textsim~0.8 on a 250 km body. However, for larger relative impactor sizes, the relationship is not so simple: While one in five 100 km parent bodies experiences at least one impact of at least one tenth of the parent body size, for both 50 km and 250 km radius bodies, that number increases to one in three; for impactors that are at least one fifth of the size of the parent body, 250 km parent bodies experience approximately twice the number of impacts than do both the 50 km and 100 km radius parent bodies. Thus, there is not a direct relationship between the number of large impactors and the size of the parent body. This is likely a consequence of the wavy nature of the size-frequency distribution; for smaller parent bodies, there are more large impactors because there are more bodies in that size range in the population: For example, there are \textsim~40 times more 5 km radius impactors than 25 km radius impactors at 10 Myr. The transition between these two effects seems to be somewhere between 50 km and 250 km radius, for impactors larger than one tenth of the parent body size. Further modeling of a range of parent body sizes could constrain this point of inflection further.

\subsubsection*{Impact energy}
For parent bodies that survived 100 Myr without disruption, the average impact energy received, \Qtot, (the sum of the impact energy from all impacts on the parent body, normalized by the parent body mass) is approximately the same for all sizes of parent bodies modeled here (Table 3). However, for those bodies that are disrupted, the larger parent bodies (\rt\ = 250 km) receive a approximately twice the total specific impact energy: A larger specific impact energy is required to disrupt a larger parent body (by a factor of 1.22 for porous materials), which results in fewer of the larger parent bodies being disrupted. This is reflected in the average impact energy from the most energetic impact ((\qdqstard)$_{\max}$), which decreases monotonically with increasing parent body size.

\begin{table*}[!t]
\footnotesize
\caption{Average total impact energy, \Qtot, sustained on a parent body, for each of the simulated scenarios.}
\begin{center}
\begin{tabular}{cccccc}
\toprule
\multirow{3}{*}{Model} & \multirow{3}{*}{f $^{\rm{a}}$} & & Geometric mean & Geometric mean & Geometric \\
& & \rt $^{\rm{b}}$ & \Qtot\ (survivors) & \Qtot\ (disrupted) & mean \\
& & [km] & [erg/g] & [erg/g] & (\qdqstard)$_{\max}$ $^{\rm{c}}$ \\
\midrule
CJS & 100 & 50 & $2.7\times10^8$ & $7.0\times10^{10}$ & 0.059 \\
CJS & 100 & 100 & $2.3\times10^8$ & $6.0\times10^{10}$ & 0.017 \\
CJS & 100 & 250 & $2.5\times10^8$ & $1.2\times10^{11}$ & 0.004 \\
\midrule
EJS & 95 & 50 & $1.4\times10^8$ & $7.0\times10^{10}$ & 0.022 \\
EJS & 95 & 10 & $1.3\times10^8$ & $9.6\times10^{10}$ & 0.007 \\
EJS & 95 & 250 & $1.1\times10^8$ & $1.2\times10^{11}$ & 0.001 \\
\bottomrule
\\
\multicolumn{6}{p{10cm}}{$^{\rm{a}}$ $f$ is the ratio of the total initial mass of the planetesimal population to the final mass of planetesimals in the asteroid belt.}\\
\multicolumn{6}{p{10cm}}{$^{\rm{b}}$ \rt\ is the radius of the parent body.}\\
\multicolumn{6}{p{10cm}}{$^{\rm{c}}$ (\qdqstard)$_{\max}$ is the ratio of the maximum impact energy from a single impact to the impact energy required for catastrophic disruption.}\\

\end{tabular}
\end{center}
\end{table*}

\subsection*{Sensitivity to collision evolution model}
Equivalent simulations were performed for the three parent body sizes discussed above, this time using the second collisional evolution model parameters (EJS model). Results for these models are summarized in Tables 2 and 3. The total number of impacts received on the surface of a planetesimal in the EJS model is a factor of \textsim~2.8 fewer than the equivalent sized parent body in the CJS model. For impactors larger than a given size (0.05 \rt, 0.1 \rt\ and 0.2 \rt), the average number of impacts per parent body is a factor of 2 -- 3 fewer in all cases listed in Table 2. This is in part because the mass of the planetesimal population depletes more rapidly in the EJS simulations (Figure 1c). Furthermore, the inclinations of bodies are more rapidly excited in the EJS simulations, which makes collisions less likely: After 1 Myr, the intrinsic collision probability is always lower in the EJS simulation compared to the CJS simulation. In the EJS simulation, \mbox{$P_i = 3.4\times10^{-18}\ \rm{km}^{-2}\ \rm{yr}^{-1}$} at 1 Myr, and decreases monotonically to \mbox{$1.6 \times 10^{-18}\ \rm{km}^{-2}\ \rm{yr}^{-1}$} by 100 Myr; however, in the CJS simulation, $P_i$ remains steady between \mbox{$\sim 3.5$ and $4.2 \times 10^{-18}\ \rm{km}^{-2}\ \rm{yr}^{-1}$} until 10 Myr, before decreasing to \mbox{$1.6 \times 10^{-18}\ \rm{km}^{-2}\ \rm{yr}^{-1}$} at 100~Myr. The fraction of parent bodies disrupted in the EJS simulation is also approximately a factor of 4 fewer than in the CJS simulation; this is as a result of the eccentric orbits of Jupiter and Saturn exciting bodies more readily in the EJS simulation, thus leading to a more rapid reduction in the number of planetesimals in the Solar System, which in turn leads to fewer collisions occurring.

Table 3 compares the amount of impact energy deposited on a parent body in the CJS and EJS simulations. The sum of the impact energy from all collisions that occur on surviving bodies is a factor of \textsim~2 -- 2.5 lower in the EJS simulations than in the CJS simulations, again, reflecting the lower number of collisions. For those parent bodies that are disrupted, the average total impact energy does not change significantly between the CJS and EJS cases (because the impact energy required for disruption is unchanged between the two models, and if a disruptive impact occurs, that impact usually delivers the most energy).

As noted above, the collisional evolution models that we employ in this work produce too much collisional grinding, leading to slightly fewer bodies between 10 and 100 km in radius. This implies that we may be overestimating the number of disruptive collisions in that size range, which should thus be considered as an upper limit.

\subsection*{Influence of porosity}
A simulation was also performed in which the parent body and all impactors were assumed to have no porosity, for the case of a 100 km radius parent body with the CJS collisional evolution parameters. Therefore, the disruption criterion used the non-porous values for $B$ and $b$ \citep{Jutzi:10}, and the calculation of heating on the parent body used results from iSALE simulations with non-porous material. The number of collisions and the frequency of disruption are similar to the porous case: 7.4\% of non- porous parent bodies are disrupted in this simulation. The major difference between the non-porous simulation and the porous simulation is the fraction of the impact energy received on the parent body that is converted to heat: In surviving bodies in the non- porous simulation, 7.2 $\pm$ 2.6\% of the total impact energy is converted to heat, and 2.8 $\pm$ 3.2\% is converted to heat in disrupted bodies (in both cases, a factor of \textsim~5 less than in the porous case). This reiterates the result that the shock compaction of pore space can greatly increase collisional heating \citep{Davison:10a}.

\msection{DISCUSSION}
The results in the previous section show that collisions on meteorite parent bodies would have been common events during the first 100 Myr. Many parent bodies would have been disrupted by energetic impacts during that time. Those that survived without being disrupted likely sustained many impacts on their surfaces, processing (compacting, heating, and mixing) the outer layers. The extent to which this occurred can vary by large amounts depending on the size of the body, and the collisional evolution of the planetesimal population, but it is unlikely that many parent bodies escaped without a significant number of impacts processing their upper few kilometers.

\subsection*{Impact histories of specific parent bodies}
In this section, the model described above is applied to examine the plausibility of some of the invocations of impacts in meteorite parent body histories. In order to examine the histories of some meteorite parent bodies, data from meteorites (such as peak temperatures, timing of impacts, etc.) can be compared with data from the parent bodies simulated in the Monte Carlo model.

\subsubsection*{Timing of t = 0}
There is one point of clarification that should be considered in this analysis: The time $t = 0$ in our model is defined as the time that the gas dispersed from the disk such that it does not significantly affect the orbits of the planetesimals. In general, meteorite studies quote timings from the formation of CAIs (Ca-Al-rich inclusions---some of the oldest known materials found in meteorites). CAIs are predicted to have formed 4568.5 $\pm$ 0.5 Myr ago \citep{Bouvier:07}. These events (gas dispersion and CAI formation) likely did not occur at the same time. \citet{Ciesla:10} argued that the CAIs which define $t = 0$ for the Solar System formed right when the solar nebula stopped accreting significant amounts of mass from its parent molecular cloud; astronomical surveys suggest disks last \textsim~2 -- 5 Myr after this point \citep{Haisch:01a}. Thus the timings discussed below could change by up to \textsim~2 Myr, and thereby alter the conclusions discussed below. The effects of moving $t = 0$ by \textsim~2 Myr (i.e. assuming that the gas dispersed 2 Myr after CAI formation---therefore, $t = 0$ in the Monte Carlo model was 2 Myr after CAI formation) are discussed below. An assumption in this analysis is that collisions between the formation of the CAIs and the dispersion of the gas are considered too gentle to be accounted for in the model. The discussion below shows that while the choice of $t = 0$ has an effect on the results of the model when looking at collision effects in the very early Solar System, its influence is diminished for events that took place after \textsim~5--10 Myr. In order to fully quantify the effects of collisions in the first several million years of Solar System history, a more thorough understanding of the timing of gas dispersion and the onset of the dynamical excitation of the planetesimal population is required.

\subsubsection*{CB chondrite parent body}
The bencubbinite CB chondrites contain large, exotic metal aggregates with elemental abundances varying strongly with volatility; the origin of these aggregates is unknown. One theory suggests that an impact between a metallic planetesimal and a reduced-silicate body could have produced a vapor plume that was rich in metal and not very oxidizing \citep{Campbell:02b}, from which the metal condensed. In order to create an appropriate vapor cloud, \citet{Campbell:02b} suggested a collision of such bodies at velocities $>$ 8 -- 10 km/s occurred before 5 Myr. Chondrules are another feature of the CB parent body: The CB chondrules are relatively young compared to other chondrules, and appear to have formed in a single event, leading to speculation that they may have an impact origin \citep{Krot:05}, and that one impact formed both the chondrules and the metal aggregates. The timing of the chondrule-forming impact has been determined from $^{207}$Pb--$^{206}$Pb ages, and is predicted to have occurred between 4.3 and 7.1 Myr after CAI formation \citep{Krot:05}. It should be noted that this chondrule formation mechanism is distinct from the impact mechanism recently suggested by \citet{Asphaug:11}, which requires impacts with velocities at $<$ 100 m/s.

To test the likelihood that an impact could have formed both the metal and the chondrules in the CB chondrite parent body, we can thus look for an impact that caused vaporization. The time of the impact must be after 4.3 Myr \citep[from][]{Krot:05}, but before 5 Myr \citep[from][]{Campbell:02b}. \citet{Campbell:02b} do not offer a constraint of the size of the colliding bodies. \citet{Krot:05} suggest an impact between Moon-sized bodies, but that constraint is simply based on requiring a velocity high enough to produce vapor, as the gravity of those bodies would increase the velocity at impact. As the planetesimals in our model are able to reach collisional velocities of 8 -- 10 km/s even at smaller sizes (due to interactions with the gas giants and the planetary embryos), here we will consider vaporizing impacts on each parent body size discussed above.

In the CJS models, on average three out of every five 50 km radius parent bodies will experience a vaporizing collision (\vimp\ $>$ 8 km/s), each 100 km radius body will experience \textsim~2.5 such collisions, and each 250 km radius body will experience over 15 vaporizing collisions. The number of collisions with \vimp\ $>$ 10 km/s is approximately 4 -- 5 times fewer than with \vimp\ $>$ 8 km/s. In the EJS models, there are around 3 times more collisions with \vimp\ $>$ 8 km/s and around 10 times more collisions with \vimp\ $>$ 10 km/s than in the CJS models. While fewer collisions are expected overall in the EJS case compared to CJS, more of these high velocity vaporizing collisions occur, due to the high eccentricity of the planetesimalsÕ orbits.

Those predictions do not account for the size of the impactor, and include all vaporizing collisions with impactors greater than 150 m in radius. If we instead only consider impactors at least one twentieth of the radius of the parent body, we get a sense of the number of large-scale vaporizing collisions. In the CJS models, around one in a hundred parent bodies will see a collision with \vimp\ $>$ 8 km/s with an impactor of that size (independent of parent body size). In the EJS models, around one in twenty parent bodies see a large impact with \vimp\ $>$ 8 km/s.

If CB chondrites were formed as hypothesized by \citet{Campbell:02b} and \citet{Krot:05}, at most \textsim~10 parent bodies would have undergone evolution allowing chondrules and metallic phases to form: To analyze whether this scenario is likely to have occurred to the CB chondrite parent body, we can examine the number of parent bodies in the collisional evolution model at 100 Myr of each size that remained in the asteroid belt. In the CJS model, there were 165 parent bodies of \textsim~50 km radius (in the bins with a radius between 40 and 63 km), 41 parent bodies of \textsim~100 km radius (in the bins with a radius between 79 and 126 km), and 3 parent bodies of \textsim~250 km radius (in the bins with a radius between 199 and 315 km). In the EJS model, these numbers are 168, 43 and 3, respectively. So, in the CJS case, one or two 50 km radius parent bodies that remain in the asteroid belt would have seen a large vaporizing collision, and there is a \textsim~40\% chance that a 100 km radius body and a \textsim~3\% chance of a 250 km radius body experiencing a vaporizing collision and surviving in the asteroid belt. In the EJS case, there would be eight 50 km radius bodies, two 100 km radius bodies and a 15\% chance of a 250 km radius body surviving in the asteroid belt that match the CB chondrite criteria. By moving the definition of $t = 0$ by 2 Myr (see above), the number of vaporizing impacts decreases by a factor of \textsim~4; this does not change the overall conclusion that it is possible that a small number of parent bodies experience the type of vaporizing collision needed to produce the metal and chondrules seen in the CB chondrites and survive past 100 Myr.

\subsubsection*{CV chondrite parent body}
A model for the formation and differentiation of the CV chondrite parent body to account for the unidirectional magnetic field was developed by \citet{Elkins-Tanton:11}. In that model, parent bodies of \textsim~100 -- 300 km radius differentiated due to the decay of $^{26}$Al, leaving an unmelted, chondritic crust of \textsim~6 -- 20 km thickness on top of a magma ocean. A porosity of 25\% is assumed for the crust, close to the porosity used in the Monte Carlo model of this work. The surviving, undifferentiated crust would then serve as the source of the pristine CV chondrites. One requirement of the model is that the \textsim~6 -- 20 km thick crust must not be disturbed during the time that the parent body is being heated and the magnetic field is in place (\textsim~10 Myr), to prevent both impact foundering and breaches leading to magma eruptions. While \citet{Elkins-Tanton:11} advocate a parent body that begins at around 100 km radius and continues to grow to 300 km radius while heating, our model cannot account for changing parent body sizes. Thus, to test this model, here we will examine the histories of parent bodies with fixed radii of 100 km and 250 km in the Monte Carlo simulation described above to determine if a parent body could have escaped any impacts large enough to penetrate the crustal material. The depth of penetration, \Hpen, of a crater in a gravity-dominated crater can be approximated as \Hpen\ = 0.28 \dtr\ \citep{O'Keefe:93}.

On 100 km radius parent bodies for the CJS case, the sum of the area of all craters would be \textsim~0.8 times the surface area of the parent body (Figures 4 \& 6). The average number of collisions that penetrate to 6 km deep is \textsim~5 per parent body. Around one in four parent bodies will experience an impact that penetrates to 20 km. Therefore, while less than 1\% of 100 km radius parent bodies survive the first 10 Myr without a collision penetrating to 6 km, around 75 -- 80\% survive without a collision that penetrates to 20 km.

For 250 km parent bodies, more collisions can be expected: \textsim~22 -- 24 collisions per parent body penetrate to 6 km, and each parent body will experience \textsim~1 collision that penetrates to 20 km depth. No 250 km radius parent bodies survives 10 Myr without a collision that penetrates to 6 km, and \textsim~40 -- 45\% of bodies survived without a collision that penetrated to 20 km.

Therefore, it seems unlikely that the CV parent body could have developed in the way proposed by \citet{Elkins-Tanton:11} if the crust was only $\leq$ 6 km thick. Our Monte Carlo model therefore predicts that for CV chondrites to come from a body such as that modeled by \citet{Elkins-Tanton:11}, a thicker crust (\textsim~20 km thick) is required: This would increase the chances that the parent body could have survived for 10 Myr without experiencing a collision that would disrupt the thermal structure and therefore the magnetic field. Whether these impacts would increase the density of the chondritic crust to create a density instability or cause it to founder outright will be the subject of future work.

Above (in the `\emph{CB Chondrite parent body}' section), we showed that approximately 40 parent bodies of \textsim~100 km radius and approximately 3 parent bodies of 250 km radius would survive to 100 Myr and remain in the asteroid belt. Therefore, if a crust of 20 km was present, around 30 parent bodies of 100 km radius would have survived past 100 Myr and matched the criteria for the CV chondrite parent body from \citet{Elkins-Tanton:11}, and only 1 parent body of \textsim~250 km size would have matched these criteria.

The number of bodies that match the CV chondrite criteria is not strongly affected by moving $t = 0$ by 2 Myr. The total number of collisions that penetrated the crust within the first 8 Myr of the model (rather than the first 10 Myr) decreased by around a third, and the number of bodies that escaped without a collision increased slightly: Around 1 -- 3\% of 100 km radius bodies escaped without a collision penetrating 6 km, and around 80 -- 85\% survived without a collision penetrating 20 km. No 250 km radius parent bodies survived without a collision penetrating a 6 km crust, and approximately half of all 250 km parent bodies survived without a collision penetrating to 20 km depth. Thus, changing the timing of $t = 0$ does not change our conclusions about the \citet{Elkins-Tanton:11} model: A thick crust (\textsim~20 km) would be needed to protect the CV chondrite parent body from a collision that could cause foundering and loss of the crust.

\subsubsection*{IAB/winonaite parent body}
The IAB non-magmatic iron meteorites and the winonaite meteorites are thought to share a common parent body \citep{Benedix:00}. While the majority of heating in these bodies likely came from $^{26}$Al decay \citep[e.g.][]{Theis:13}, there are some studies of the chronology of these meteorites that suggest that the parent body experienced several heating events after the decay of short-lived radionuclides. Impacts have been invoked to explain these relatively late thermal events \citep{Schulz:09,Schulz:10,Schulz:12} which are recorded by the Hf-W chronometer as well as Sm and W isotopic abundances. An event at $2.5^{+2.3}/_{-2.0}$ Myr caused silicate differentiation; partial melting was caused at 5.06$^{+0.42}/_{-0.41}$ Myr; further thermal metamorphism occurred at 10.8$^{+2.4}/_{-2.0}$ Myr with temperatures reaching 1273 K; and finally a disruptive event occurred after \textsim~12 Myr. The extent to which impacts could provide the heat required to match the thermal history recorded in meteorites can be evaluated using the Monte Carlo model. Figure 10 shows the probability that a given fraction of a 100 km radius parent body (in the CJS simulation) is heated to temperatures above those of chondrites with petrologic type 6 ($>$ 1273 K) or above the assumed (dunite) solidus ($>$ 1373 K) in any impacts that occurred after 10 Myr (i.e. after the decay of $^{26}$Al). In most parent bodies, only a small fraction is heated to those temperatures: Less than one hundredth of the parent body reaches these temperatures in more than half of all cases modeled. However, in some bodies, a significant fraction of the parent body is heated in late collisions: In 7.5\% of parent bodies, more than one tenth of the parent body is heated to the required temperatures. The Monte Carlo model can also be used to examine the impact history of this parent body in more detail. Of the parent bodies that were disrupted after 12 Myr, each collision was examined to determine if it heated material to the required temperature in the given time intervals. In the simulation of 100 km radius parent bodies with the CJS collisional evolution model, 3834 parent bodies (out of 105) were disrupted after 12 Myr. Of those bodies, only 3 (around 0.1\% of the disrupted bodies) had energetic enough collisions during each of the three given time periods listed above, to heat at least one thousandth of the parent body to the required temperature. It is unclear if the first heating event (at 2.5$^{+2.3}/_{-2.0}$ Myr) was caused by impacts or radionuclide decay \citep{Schulz:10}, as it is during the time when $^{26}$Al decay was still active. Running the same analysis, but without the requirement for an impact to heat material in that event yields 54 parent bodies that match the criteria (around 1.4\% of the disrupted bodies). For the EJS model, 894 parent bodies (out of 105) were disrupted after 12 Myr; of those, 8 parent bodies (\textsim~1\% of the disrupted bodies) experienced collisions to match all the events, and 48 parent bodies (\textsim~5\%) met the criteria when the first heating event was excluded. This shows that the collisional history needed to produce the IAB/winonaite parent body may be difficult to achieve.

\begin{figure}[!t]
\includegraphics[width=0.9\linewidth]{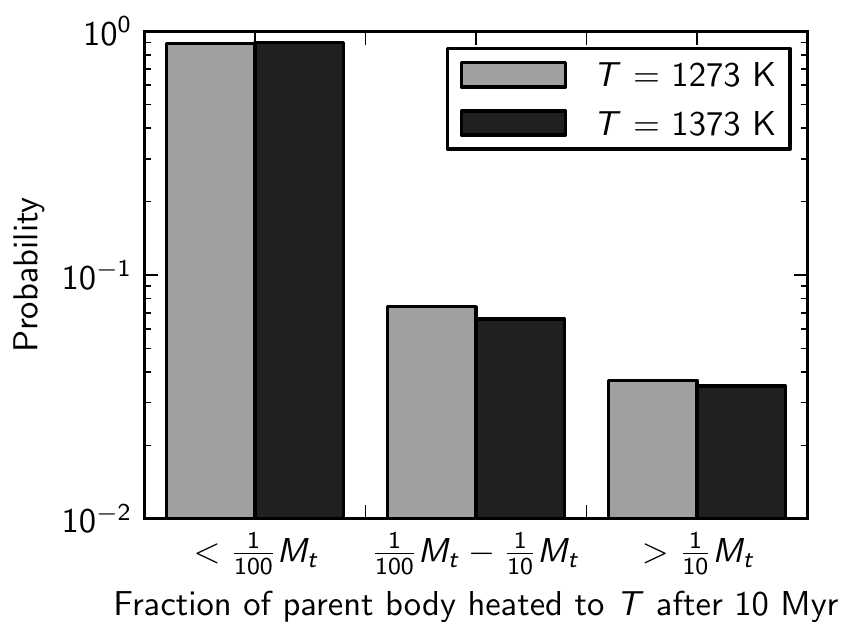}
\caption{Probability of a given fraction of a 100 km radius parent body in the CJS model being heated above 1273 K (i.e. above petrologic type 6) or 1373 K (i.e. above the dunite solidus) in impacts that occurred later than 10 Myr after formation. In almost 90\% of all parent bodies in this simulation, less than one hundredth of the parent body was heated to either of those temperatures. In 7\% of cases, between one hundredth and one tenth of the parent body was heated to those temperatures, and in around 3.5\% of cases, more than one tenth of the parent body was heated to those temperatures.}
\end{figure}

To evaluate the influence of our choice of $t = 0$ on the analysis of the IAB/winonaite parent body, the times of the impact events were shifted by 2 Myr. In the CJS model, the total number of disrupted parent bodies that could match the criteria decreased from 54 to 11, and in the EJS model that number decreased slightly from 42 to 37. Thus, the choice of $t = 0$ does not significantly change our conclusions about the IAB/winonaite parent body.

A caveat to the approach used here is that it is assumed that the material starts out cold before the impact, and therefore requires a more energetic collision to produce the levels of heating observed in the meteorite record; this approach does not take into account the heating provided by short-lived radionuclides, and therefore should be taken as a lower estimate of the probability of producing a parent body to match the IAB/winonaite body. Furthermore, here we have assumed a parent body radius of 100 km---however, the size of the IAB/winonaite parent body is not well constrained. A different parent body size could yield different results. Further work is clearly needed to fully quantify the early impact history of the IAB/winonaite parent body.

\subsubsection*{H chondrite parent body}
Models of the thermal evolution of the H chondrite parent body predict an onion shell structure resulting from the heat produced by the decay of short-lived radionuclides, such as $^{26}$Al \citep[e.g.][]{Trieloff:03, Kleine:08, Harrison:10a}. However, \citet{Harrison:10a}, in order to explain some measurements of cooling rates, peak temperatures and closure times which do not fit their onion shell model, invoked an impact which disturbed the thermal evolution of a portion of the body by excavating material from at least 5.6 km deep on a 100 km radius parent body. In the simulations presented above for a 100 km parent body, on average 2.3 collisions per parent body excavated to this depth in the CJS model (see Figure 5b), and in the EJS model, 1.5 collisions per parent body matched this criterion. Approximately 8.4\% and 22.7\% of parent bodies modeled escaped without any collisions excavating to this depth in the CJS and EJS simulations, respectively. While the results depend strongly on the choice of collisional evolution model (CJS or EJS), they show that most parent bodies (more than 77\% of all 100 km radius parent bodies) will experience at least one impact of the type required by \citet{Harrison:10a}. Of the 40 or so \textsim~100 km radius parent bodies that survived to 100 Myr, at least 30 would have experienced a collision which excavated to at least 5.6 km depth. Thus, the requirement of an impact disturbing the H chondrite parent body thermal structure is supported by this study.

Other models of the H chondrite parent body suggest impacts may have played a larger role in the thermal history \citep{Scott:11a}, excavating and thoroughly mixing material \citep{Krot:12}. In that model, both type 4 and type 6 materials are brought to the surface by impacts early in the evolution of the parent body. The depth to type 4 and type 6 material can be estimated using an onion shell model. \citet{Harrison:10a} predict type 4 material may be as shallow as 0.9 km below the surface of a 100 km radius parent body. Analysis of cooling rates from one type 6 meteorite, predicted type 6 material at 3.4 km depth. However, other analyses in that study using both cooling rates and closure times predicted a depth of at least 11.2 km for type 6 meteorites. In the CJS model of a 100 km radius parent body, \textsim~123 impacts per parent body excavated material from 0.9 km depth, \textsim~9 impacts excavated from a depth of 3.4 km and \textsim~0.3 impacts per parent body excavated from 11.2 km. In the EJS model, fewer impacts excavated to those depths with \textsim~56 impacts excavated from 0.9 km, \textsim~5 impacts excavated from 3.4 km, and \textsim~0.2 impacts excavated from 11.2 km. This shows that bringing type 4 material to the surface is a common process, and should occur in at least 50 events per parent body in the first 100 Myr. In the CJS model, the surface area of craters that excavated to this depth is three quarters of the area of the parent body; in the EJS model, on average around a third of the parent body surface would experience craters that excavate to this depth. If type 6 material is as shallow as 3.4 km, up to 9 events could bring that material to the surface to cool more quickly (more than half of the surface area is affected in this way in the CJS model; one quarter in the EJS model). However, if type 6 material is deeper, then far fewer impacts would be able to excavate it---not every parent body is likely to experience a collision during the first 100 Myr of the magnitude required to excavate material from a depth of 11.2 km (an average covering of two fifths of the parent body surface area in the CJS model and one fifth in the EJS model).

\msection{CONCLUSIONS}
In this study, we have developed a model to simulate the collisional histories of meteorite parent bodies. The model predicts that impacts were common processes on parent bodies and the outer layers of those bodies would have been strongly processed by shock events. Impacts can provide a significant, secondary heat source to parent bodies, which can be especially important after the decay of short-lived radionuclides and in the regions near the surface of the original parent body. The probabilities of different outcomes vary strongly with the different dynamical scenarios considered for the early Solar System. Further compilation of how shock processes affected meteorite parent bodies in the early Solar System should allow us to test the different dynamical scenarios and collisional histories envisioned for terrestrial planet formation.

The model has been applied to evaluate the proposed histories of several meteorite parent bodies: At most, around 10 parent bodies that survived to 100 Myr underwent a collisional evolution consistent with the formation mechanism of the chondrules and metallic phases of \citet{Campbell:02b} and \citet{Krot:05}; around 1 -- 5\% of bodies that were disrupted after 12 Myr (which comprises \textsim~1 -- 4\% of the initial population of 100 km planetesimals) experienced impact events contemporaneously with some late heating events on the IAB/winonoaite parent body \citep{Schulz:09, Schulz:10}; at least 30 of the \textsim~40 surviving 100 km radius bodies would sustain an impact that excavated material from a depth of at least 5.6 km, as predicted by \citet{Harrison:10a} for the H chondrite parent body, and almost all bodies experienced impacts that brought type 4 material to the surface, as required by \citet{Scott:11a}; a thick crust (\textsim~20 km) is required on the CV parent body to prevent foundering and disturbance of the thermal structure---a crust of this thickness would provide enough shielding for 10 Myr for \textsim~30 parent bodies to survive for 100 Myr and fit the model of \citet{Elkins-Tanton:11}, while a thin crust (\textsim~6 km) would experience too many impacts that puncture it and thus allow foundering.

There are some critical uncertainties in the Monte Carlo simulations which should be addressed in future iterations of the model: For example, the initial mass of material in the planetesimal population, the initial orbits of the gas giants (CJS or EJS), the response of the disruption threshold to temperature and porosity changes, and the evolution of the porosity structure of a parent body through time are all factors for which we have chosen the best available approximations. Nonetheless, our approach offers a new means for evaluating the roles that impacts played in shaping meteorite parent bodies. By better recognizing the signatures and timing of impact events in meteorites, we can understand the collisional history of meteorite parent bodies and use that information to constrain dynamical models for the Solar System.

{\it 
Acknowledgments: We gratefully acknowledge the major contributions of Jay Melosh, Boris Ivanov, Kai WŸnnemann and Dirk Elbeshausen to the development of iSALE. We thank Ed Scott, Bill Bottke and Robert Grimm for their insightful and constructive comments. TMD and FJC were supported by NASA PGG grants NNX09AG13G and NNX12AQ06. TMD and GSC were supported by STFC grant ST/J001260/1. GSC was also supported by STFC grant ST/G002452/1. DPO was supported by grant NASA PGG grant NNX09AE36G.
}

\appendix
\renewcommand\thefigure{\thesection A\arabic{figure}}    
\setcounter{figure}{0}    

\msection{APPENDIX A: VFD PARAMETERISATION}
In order to use the results from the dynamical and collisional evolution simulations, they must be parameterised in such a way that the Monte Carlo model can use them. In Figure 1, the mean velocity is presented as a function of time. In order to allow a VFD to be produced at every timestep in the Monte Carlo model, the mean velocity is interpolated, and converted into a VFD using a Maxwellian distribution. To check that a Maxwellian distribution is appropriate, the VFD from the dynamical models was output at times throughout the simulation, and compared to the Maxwellian. The dynamical model output is well fit by the Maxwellian distribution (with a coefficient of determination, $R^2 > 0.96$ for all timesteps in the first 100 Myr). The evolution of the VFD for the CJS simulation is shown in Figure A1.

\begin{figure}[!t]
\includegraphics[width=0.9\linewidth]{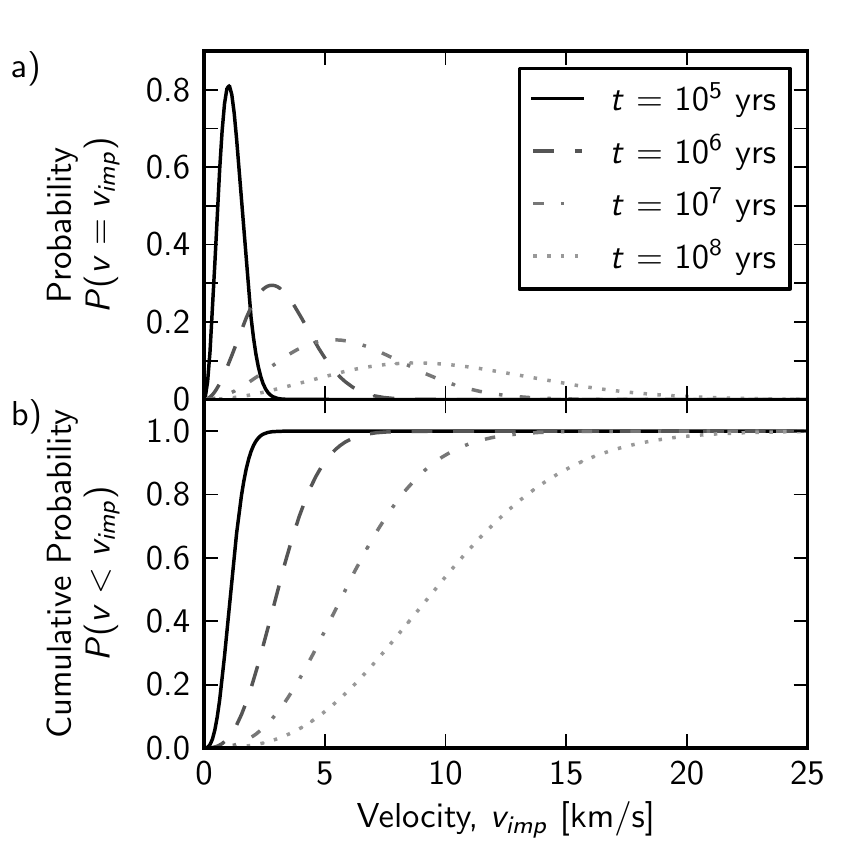}
\caption{Evolution of the VFD for the CJS simulation, at several points in time. The VFD is fit well (R2 > 0.96) by a Maxwellian distribution.}
\end{figure}

\bibliographystyle{apalike}
\bibliography{library}
\balance
\end{document}